\documentclass[12pt,a4paper]{article}
\usepackage{graphics}
\usepackage{axodraw}

\newcommand\Rsep{R_{\mathrm{sep}}}
\newcommand\Rmax{{R_{\mathrm{max}}}}
\newcommand\ptmin{p_{t\mathrm{min}}}
\newcommand\ptjet{p_{t\mathrm{jet}}}
\newcommand\gE{{\gamma_{\scriptscriptstyle E}}}
\newcommand\as{\alpha_{\mathrm{s}}}
\newcommand\ab{{\bar{\alpha}_p}}
\newcommand\abz{{\bar{\alpha}_0}}
\newcommand\kt{{k_\perp}}
\newcommand\V{{\cal{V}}}
\newcommand\smfrac[2]{{\textstyle\frac{#1}{#2}}}
\newcommand{\ltap}{\;\raisebox{-.4ex}{\rlap{$\sim$}} \raisebox{.4ex}{$<$}\;}
\newcommand{\gtap}{\;\raisebox{-.4ex}{\rlap{$\sim$}} \raisebox{.4ex}{$>$}\;}

\begin{document}

\begin{titlepage}
\begin{flushright}
  RAL-97-026 \\ hep-ph/9707338
\end{flushright}
\par \vspace{10mm}
\begin{center}
{\Large \bf
Jet Shapes in Hadron Collisions:\\
Higher Orders, Resummation\\[1ex] and Hadronization}
\end{center}
\par \vspace{2mm}
\begin{center}
{\bf M.H. Seymour}\\
\vspace{5mm}
{Rutherford Appleton Laboratory, Chilton,}\\
{Didcot, Oxfordshire.  OX11 0QX.  U.K.}
\end{center}

\par \vspace{2mm}
\begin{center} {\large \bf Abstract} \end{center}
\begin{quote}
\pretolerance 10000
The jet shape is a simple measure of how widely a jet's energy is
spread.  At present jet shape distributions have only been calculated to
leading order in perturbative QCD.  In this paper we consider how much
these predictions should be affected by higher order perturbative
corrections, by resummation of enhanced corrections to all orders, and
by (power-suppressed) non-perturbative corrections.  We also show that
current cone-type jet definitions are not infrared safe for final states
with more than three partons.  Unless this situation is rectified by
using improved definitions, hadron collider experiments will never be
able to study the internal properties of jets with the quantitative
accuracy already achieved in $e^+e^-$ annihilation.
\end{quote}

\vspace*{\fill}
\begin{flushleft}
  RAL-97-026 \\ July 1997
\end{flushleft}
\end{titlepage}

\section{Introduction}

Hadron collisions at the Tevatron are a prolific source of high
transverse momentum jets.  The collider experiments, CDF and D\O, have
made very high precision measurements of the inclusive jet
rates[\ref{CDF1},\ref{D01}], as well as the rate of $n$-jet events,
$n\ge2$[\ref{CDF2},\ref{D02}].  At present, perturbative calculations
extend to next-to-leading order (NLO) for the
inclusive[\ref{ACGG},\ref{EKS1},\ref{GGK1}] and
dijet[\ref{EKS2},\ref{GGK2}] rates and the NLO three-jet rate is close
to completion[\ref{GK}].  These allow quite precise studies of QCD,
albeit for very inclusive quantities.

At NLO, the cross sections begin to depend on the exact definition of
the jets, since the jets begin to develop internal structure.  The
experiments have used measures of this internal structure as a
cross-check on the reliability of the calculations, as well as as a
study of QCD in its own right.  In fact, as we shall argue, these more
exclusive event properties contain considerably more information about
QCD dynamics, and make an ideal place to study QCD.

One of the most popular measures of jet structure in hadron collisions
is the `jet shape': the fraction of the jet's energy within a cone of a
given size, centred on the jet direction.  This has been measured by
both CDF and D\O\ over wide ranges of phase space[\ref{CDF3},\ref{D03}].
It has also been measured in photoproduction by the ZEUS
collaboration[\ref{ZEUS}] with essentially the same method, so in this
context we can class photoproduction as a hadron collision.  Making
these studies in photoproduction has several advantages that offset the
fact that they cannot reach as high jet~$E_T$.  The direct component of
the photon means that events can be selected with very little underlying
event activity.  Once these are understood, one can change the kinematic
cuts to bring in the resolved component and thereby study the underlying
event contribution in detail.  Similarly, by varying the event
kinematics, one can vary the quark/gluon jet mix in a way that is under
much better perturbative control than in hadron collisions.

The NLO prediction of jet rates gives a leading order (LO) prediction of
the jets' internal structure[\ref{EKS3}].  Thus the calculations with
which the experiments have so far compared are not expected to be
terribly accurate.  In this paper we discuss the extent to which this
could be improved by including higher order terms in perturbation
theory, by summing terms to all orders, and by including an account of
hadronization.  The main conclusion is that these effects together
constitute a very large effect on the jet shape, and we should not be
surprised if the LO calculations do not give very good fits to data.

As part of this study, we considered how reliable the jet definitions
used by most experiments are for these kinds of study.  The conclusion
is that they are not very reliable.  In order to make quantitative
comparisons with theory beyond LO, this situation must be improved,
either by using a cluster-type algorithm, or by a simple modification of
existing cone-type algorithms.  Because of this, we have not compared
our results with existing experimental measurements.  Our emphasis is
instead on understanding what physical effects are important in
determining the jet shape, how large various terms are, and how accurate
a future, more complete, calculation could be.

We begin in Sect.~2 by reviewing the jet definitions in current use.  In
a simple approximation, we make an order by order expansion, and show
that calculations for current cone-based definitions will break down at
next-to-next-to-leading order (NNLO) for jet cross sections (and
therefore at NLO for internal jet properties).  We discuss a simple
remedy, and show that this has a well-defined perturbative expansion.

Section~3 contains the real substance of the paper, in which we
calculate the jet shape in several steps, with ever-increasing accuracy.
We start at LO, then add higher orders in perturbation theory, followed
by running coupling effects, all-orders resummation of enhanced terms
and (power-suppressed) hadronization corrections.

A recent paper[\ref{GGK3}] proposed using the first moment of the jet
shape as a probe of the jet shape that is less susceptible to
non-perturbative corrections.  In Sect.~4 we consider this in the light
of our studies, and find that it is not significantly better than the
full distribution.

Finally in Sect.~5 we discuss the implications of our results and the
outlook for the future.

\section{Jet definitions, safeness and jet cross sections}

By now, collider experiments are used to comparing their data with NLO
calculations.  As a result, all currently-used jet definitions are
infrared and collinear safe to at least NLO, i.e.~when there can be up to
two partons in a jet.  However, this is not sufficient to guarantee a
well-defined perturbative expansion, because unsafe behaviour can still
arise at higher orders, i.e.~when there can be many partons in a jet.

Up to now, these issues have not had to be discussed in detail.  This is
because the data have always been compared either with NLO calculations, or
with hadron-level Monte Carlo event generators.  In the former, the jet
definitions are infrared and collinear safe by construction.  In the
latter, the typical hadronic scale $\sim\!1$~GeV cuts off all divergent
integrals, yielding finite, but fundamentally non-perturbative, cross
sections.

Since in this paper we aim to calculate, at least approximately, the
effect of high perturbative orders, we are forced to tackle these issues
head-on.  In this section, we discuss the two main jet algorithms in
current use: the $\kt$ algorithm[\ref{CDSW}] and the iterative cone
algorithm[\ref{CDF4}].  While the first is safe to all orders in
perturbation theory, we shall see that the second falls into a class
that we call `almost unsafe', to be defined shortly.  We show that a
simple modification, proposed by Steve Ellis many years ago[\ref{LdP}],
renders it safe to all orders.

Since each experiment has defined their own slightly different version
of the iterative cone algorithm, we concentrate on just one of them, the
D\O\ algorithm\footnote{We are grateful to members of the D\O\ 
  collaboration for making full details of their algorithm available to
  us[\ref{D04}].}, and only indicate the differences with respect to
other versions where relevant.

\subsection{Jet definitions}

All the algorithms we discuss define the momentum of a jet in terms of
the momenta of its constituent particles in the same way, inspired by
the Snowmass accord[\ref{Snowmass}].  The transverse energy, $E_T$,
pseudorapidity, $\eta$, and azimuth, $\phi$, are given by:
\begin{eqnarray}
\label{snowmass}
  E_{T\mathrm{jet}} &=& \sum_{i\in\mathrm{jet}} E_{Ti},\nonumber\\
  \eta_{\mathrm{jet}} &=& \sum_{i\in\mathrm{jet}}
  E_{Ti}\,\eta_{i}/E_{T\mathrm{jet}},\\
  \phi_{\mathrm{jet}} &=& \sum_{i\in\mathrm{jet}}
  E_{Ti}\,\phi_{i}/E_{T\mathrm{jet}}.\nonumber
\end{eqnarray}
We shall always use boost-invariant variables, so whenever we say
`angle', we mean the Lorentz-invariant opening angle $R_{ij} =
\sqrt{(\eta_i-\eta_j)^2 + (\phi_i-\phi_j)^2}$.  Also, whenever we say
`energy', we mean transverse energy, $E_T=E\sin\theta$.

\subsubsection{The \boldmath$\kt$ algorithm}

We discuss the fully-inclusive $\kt$ algorithm including an $R$
parameter[\ref{ES}].  It clusters particles (partons or calorimeter
cells) according to the following iterative steps:
\begin{enumerate}
\item
  For every pair of particles, define a closeness
\begin{equation}
  d_{ij} = \min(E_{Ti},E_{Tj})^2R_{ij}^2
  \left(\approx \min(E_i,E_j)^2\theta_{ij}^2 \approx k_\perp^2\right).
\end{equation}
\item
  For every particle, define a closeness to the beam particles,
\begin{equation}
  d_{ib} = E_{Ti}^2 R^2.
\end{equation}
\item
  If $\min\{d_{ij}\}<\min\{d_{ib}\},$ {\em merge\/} particles $i$ and
  $j$ according to Eq.~(\ref{snowmass}) (other merging schemes are also
  possible[\ref{CDSW}]).
\item
  If $\min\{d_{ib}\}<\min\{d_{ij}\},$ jet $i$ is {\em complete}.
\end{enumerate}
These steps are iterated until all jets are complete.  In this case, all
opening angles within each jet are $<R$ and all opening angles between
jets are $>R$.

\subsubsection{The D\O\ algorithm}

Since this is the main algorithm we shall study, we define it in full
detail.  It is based on the iterative-cone concept, with cone radius
$R$.  Particles are clustered into jets according to the following
steps:
\begin{enumerate}
\item The particles are passed through a calorimeter with cell size
  $\delta_0\times\delta_0$ in $\eta\times\phi$ (in D\O, $\delta_0=0.1$).
  In the parton-level algorithm, we simulate this by clustering together
  all partons within an angle $\delta_0$ of each other.
\item Every calorimeter cell (cluster) with energy above $E_0$, is
  considered as a `seed cell' for the following step (in D\O,
  $E_0=1$~GeV).
\item\label{reiterate} A jet is defined by summing all cells within an
  angle $R$ of the seed cell according to Eq.~(\ref{snowmass}).
\item If the jet direction does not coincide with the seed cell,
  step~\ref{reiterate} is reiterated, replacing the seed cell by the
  current jet direction, until a stable jet direction is achieved.
\item We now have a long list of jets, one for each seed cell.  Many are
  duplicates: these are thrown away\footnote{In D\O, any with energy
    below 8~GeV are also thrown away.  For jets above 16~GeV, this makes
    only a small numerical difference, which is not important to our
    discussion, so we keep them.}.
\item\label{merge} Some jets could be overlapping.  Any jet that has
  more than 50\% of its energy in common with a higher-energy jet is
  merged with that jet: all the cells in the lower-energy jet are
  considered part of the higher-energy jet, whose direction is again
  recalculated according to Eq.~(\ref{snowmass}).
\item\label{split} Any jet that has less than 50\% of its energy in
  common with a higher-energy jet is split from that jet: each cell is
  considered part only of the jet to which it is nearest.
\end{enumerate}
Note that despite the use of a fixed cone of radius $R$, jets can
contain energy at angles greater than $R$ from their direction, because
of step~\ref{merge}.  This is not a particular problem.  This is
essentially also the algorithm used by ZEUS (PUCELL), except that their
merging/splitting threshold is 75\% instead of 50\%.

\subsubsection{The CDF algorithm}

The CDF algorithm is rather similar to D\O's, but with a different
splitting procedure:
\begin{enumerate}
\item[\ref{merge}.] Any jet that has more than 75\% of its energy in
  common with a higher-energy jet is merged with that jet: all the cells
  in the lower-energy jet are considered part of the higher-energy jet,
  whose direction is again recalculated according to
  Eq.~(\ref{snowmass}).
\item[\ref{split}.] Any jet that has less than 75\% of its energy in
  common with a higher-energy jet is split from that jet: each cell is
  considered part only of the jet to which it is nearest.  The
  directions of the two jets are then recalculated by iterating
  step~\ref{reiterate}.
\end{enumerate}

\subsection{Infrared and collinear safeness}

The formal definition of infrared safeness is as follows:
\begin{equation}
  \parbox{12cm}{\em An observable is infrared safe if, for \emph{any}
  $n$-parton configuration, adding an infinitely soft parton does not
  affect the observable at all.}
\end{equation}
The formal definition of collinear safeness is similar:
\begin{equation}
  \parbox{12cm}{\em An observable is collinear safe if, for \emph{any}
  $n$-parton configuration, replacing any massless parton by an exactly
  collinear pair of massless partons does not affect the observable at
  all.}
\end{equation}
We distinguish three types of observable:
\begin{itemize}
\item[]\emph{Safe}: Both the above conditions are fulfilled.  Cross
  sections are calculable order by order in perturbation theory, with
  hadronization effects resulting in only power-suppressed corrections.
\item[]\emph{Unsafe}: One or both the conditions is violated.  Cross
  sections are infinite in perturbation theory, signalling that they are
  fundamentally non-perturbative.
\item[]\emph{Almost unsafe}: At first sight it appears that one or both the
  conditions is violated, but in fact close inspection of the algorithm
  shows that seemingly-minor details actually rescue its safeness.  Cross
  sections are calculable order by order in perturbation theory, but can be
  rather unreliable, owing to the reliance on (typically) small parameters.
\end{itemize}

Unsafe or almost unsafe algorithms often seem perfectly safe up to a
certain order in perturbation theory, after which their unsafeness
reveals itself.  This is the case in the iterative cone algorithm, in
both its D\O\ and CDF variants, which are almost unsafe, despite being
fully safe at NLO.  As we shall shortly show, it is the fact that
the particles have been passed through a calorimeter, with finite cell
size and minimum energy trigger, that renders it finite.  One might
naively think that these minor details should make small contributions
(vanishing as $E_0\to0$ or $\delta_0\to0$).  However, we shall see that
the energy threshold, $E_0$, becomes logarithmically important at NNLO,
and we also argue that the cell size, $\delta_0$, will at NNNLO.  This
can be dangerous for (at least) three reasons: firstly, it means that
anyone making a calculation must know the very precise definitions of
how the data are treated~-- what size and shape the calorimeter cells
are, what energy they trigger at, whether they are preclustered, what
preprocessing is done, etc, and must implement them in the parton-level
calculation; secondly, since $E_0$ and $\delta_0$ are rather small,
higher-order perturbative corrections are expected to be large, as are
(power-suppressed) hadronization corrections, rendering the resulting
calculations unreliable; and finally in some cases, the features that
render the observable finite seem so minor that they are not reported at
all or considered part of the experimental set-up and `corrected for',
i.e.~an almost unsafe definition is `corrected' to an unsafe one.

This problem was first encountered by Kilgore and Giele[\ref{GK}].  In
their calculation of three jets at NLO, they ignored details of the
calorimeter, effectively setting $E_0=\delta_0=0$, and encountered an
infrared divergence arising from the separation of three nearby partons,
one of which is arbitrarily soft, into one or two jets.  It turned out
that in the case of the three-jet analysis they are concerned with, this
situation was rendered `almost unsafe' by an additional cut imposed by
CDF~-- the jets should be well separated.  Once this is applied, the jet
cross section can be calculated order by order in perturbation theory.
To our knowledge, no such cut is made in the inclusive jet cross section
measurement, so unless one explicitly includes calorimeter information
in the calculation it is genuinely unsafe, which would first manifest
itself at NNLO.

\subsection{Jet cross sections}
\label{secxsec}

As a concrete example of the above considerations, we calculate the
inclusive jet cross section up to NNLO accuracy (up to three partons per
jet).  We use an extremely simple model, which is certainly too simple for
quantitative predictions, but which is sufficient to illustrate the
problems of infrared and collinear safeness, since it is based on the soft
and collinear approximations.  We consider a gluon-only world, in which the
LO inclusive gluon production cross section is
\begin{equation}
  \frac{d\sigma_0}{dp_t\,d\eta} = \frac{A^{n-3}}{p_t^n},
\end{equation}
where $A$ is an arbitrary parameter with dimensions of energy, and $n$ is
of order 5 for typical $p_t$ values of interest.  We consider the
double-logarithmic approximation (DLA), but with some modifications:
energy-momentum is conserved at vertices according to Eq.~(\ref{snowmass});
strict angular ordering is imposed.  The former is two logarithms down on
the DLA, while the latter is only one, but both are essential for making
simple numerical treatments.  In this approximation, the coupling does
not run.

We also simplify the kinematics, by assuming that the $\eta\times\phi$
plane is infinite in all directions.  Soft radiation is limited to a cone
of size $\Rmax$.  For concreteness, we set $\Rmax=2,$ although any size
considerably greater than $R$ would give similar results.

For the sake of having a simple quantity to calculate, we consider the
fully inclusive jet cross section:
\begin{equation}
  \sigma(\ptmin) \equiv \int_{\ptmin}^\infty dp_t \;
  \left.\frac{d\sigma}{dp_t\,d\eta}\right|_{\eta=0}.
\end{equation}
Clearly the LO cross section is
$\sigma_0=\frac1{n-1}A^{n-3}/\ptmin^{n-1}$.  In all numerical results we
set $\ptmin=50$~GeV, although in this simple model, the dependence of
the higher order corrections on $\ptmin$ is extremely weak.

At NLO, the DLA real contribution is:
\begin{equation}
  d\sigma_{1R} =
  \sigma_0\frac{2C_A\as}{\pi} \; \frac{dr}{r} \Theta(\Rmax-r) \;
  \frac{dz}{z} \; \frac{d\varphi}{2\pi},
\end{equation}
and the virtual contribution is:
\begin{equation}
  \sigma_{1V} =
  -\sigma_0\frac{2C_A\as}{\pi}
  \left(\int_0^\Rmax \frac{dr}{r} \int_0^1
  \frac{dz}{z} - \V\right).
\end{equation}
Each is separately divergent, such that the sum of the two is finite for
any infrared safe jet definition.  Note that we have inserted a
completely arbitrary finite virtual term to emphasize the arbitrariness
of the overall normalization.  In the numerical plots we set $\V=1$,
which roughly reproduces the normalization of the full NLO calculation.

For any of the definitions above, the results are the same:
\begin{eqnarray}
\sigma_1 &=& \frac{2C_A\as}{\pi} \int_{\ptmin}^\infty d\ptjet
\; \int_0^\infty dp_t \frac{A^{n-3}}{p_t^n} \; \Biggl(
\int_R^\Rmax \frac{dr}{r} \;
\int_0^1 \frac{dz}{z} \nonumber\\&&\;\; \Bigl\{ \delta(\ptjet-zp_t) +
\delta(\ptjet-(1-z)p_t) - \delta(\ptjet-p_t) \Bigr\}
+\V \; \delta(\ptjet-p_t)\Biggr).
\phantom{(99)}
\end{eqnarray}
The first two terms in curly brackets are contributions where the two
real partons are separate jets, while the third is part of the virtual
contribution~-- the rest has exactly cancelled the real emission
contribution in which the two partons are in the same jet.  Clearly the
last two terms cancel as $z\to0$, and the whole thing is finite.
Integrating out $p_t$ on the $\delta$-functions, we obtain:
\begin{eqnarray} \sigma_1 &=& \frac{2C_A\as}{\pi}
  \int_{\ptmin}^\infty d\ptjet \frac{A^{n-3}}{\ptjet^n} \;
  \left(
    \int_R^\Rmax \frac{dr}{r} \; \int_0^1 \frac{dz}{z} \Bigl\{ z^{n-1} +
    (1-z)^{n-1} - 1 \Bigr\}
  +\V\right)
\phantom{(99)}
  \\
  &=& -\sigma_0 \; \frac{2C_A\as}{\pi} \;
  \left(
    \log\frac\Rmax R \Bigl\{ \psi(n-1)+\gE \Bigr\}
  -\V\right).
\end{eqnarray}
For $n=5$, the curly bracket gives $11/6$, although of course this is
nothing to do with the usual collinear $11/6$.

We can see that the DLA is not strictly a good enough approximation of
this cross section, since all $z$ values contribute equally to the final
result, yielding only a single logarithm (except for large~$n$, where
$\psi(n-1)\sim\log n$).  Thus we should include hard gluon corrections
and gluon splitting to quark-antiquark pairs.  However, it is perfectly
sufficient for our illustrative purposes.

Moving to NNLO, the double-real emission cross section is given by:
\begin{eqnarray}
  d\sigma_{2(2)} &=& \sigma_0\left(\frac{2C_A\as}{\pi}\right)^2
  \; \frac{dr_1}{r_1} \Theta(\Rmax-r_1) \; \frac{dz_1}{z_1} \;
  \frac{d\varphi_1}{2\pi}
  \; \frac{dr_2}{r_2} \Theta(r_1-r_2) \; \frac{dz_2}{z_2} \;
  \frac{d\varphi_2}{2\pi}
\nonumber\\
  &+& \sigma_0\left(\frac{2C_A\as}{\pi}\right)^2
  \; \frac{dr_1}{r_1} \Theta(\Rmax-r_1) \; \frac{dz_1}{z_1} \;
  \frac{d\varphi_1}{2\pi}
  \; \frac{dr_2}{r_2} \Theta(r_1-r_2) \; \frac{dz_2}{z_2} \;
  \frac{d\varphi_2}{2\pi}.
\end{eqnarray}
Although the two terms look identical, they correspond to configurations
where the second gluon is emitted from the hard (first term) or soft
(second term) gluon from the first splitting (i.e.~they would have
different colour factors if the initiating parton was a quark, $C_F^2$ and
$C_FC_A$ respectively).  The kinematics have to be calculated accordingly.

The single-real emission cross section is given by:
\begin{eqnarray}
  d\sigma_{2(1)} &=& -\sigma_0\left(\frac{2C_A\as}{\pi}\right)^2
  \; \frac{dr_1}{r_1} \Theta(\Rmax-r_1) \; \frac{dz_1}{z_1} \;
  \frac{d\varphi_1}{2\pi}
  \left(
  \int_0^\Rmax \frac{dr_2}{r_2} \int_0^1 \frac{dz_2}{z_2}
  -\V\right)
\nonumber\\
  && -\sigma_0\left(\frac{2C_A\as}{\pi}\right)^2
  \; \frac{dr_1}{r_1} \Theta(\Rmax-r_1) \; \frac{dz_1}{z_1} \;
  \frac{d\varphi_1}{2\pi}
  \int_0^{r_1} \frac{dr_2}{r_2} \int_0^1 \frac{dz_2}{z_2},
\end{eqnarray}
where the two terms are of the `$C_F^2$' and `$C_FC_A$' type
respectively.  Note that we do not insert a finite virtual term in the
`$C_FC_A$'-type term.  Such a term could be absorbed into the running
coupling, which we do not keep track of here.

Finally, the no-emission cross section is given by:
\begin{equation}
  d\sigma_{2(0)} = \sigma_0\left(\frac{2C_A\as}{\pi}\right)^2
  \;\frac12\;
  \left(\int_0^\Rmax \frac{dr_1}{r_1} \int_0^1 \frac{dz_1}{z_1}-\V\right)
  \left(\int_0^\Rmax \frac{dr_2}{r_2} \int_0^1 \frac{dz_2}{z_2}-\V\right),
\end{equation}
which is of the `$C_F^2$' type only.

\begin{figure}
  \centerline{
    \resizebox{!}{6cm}{\includegraphics{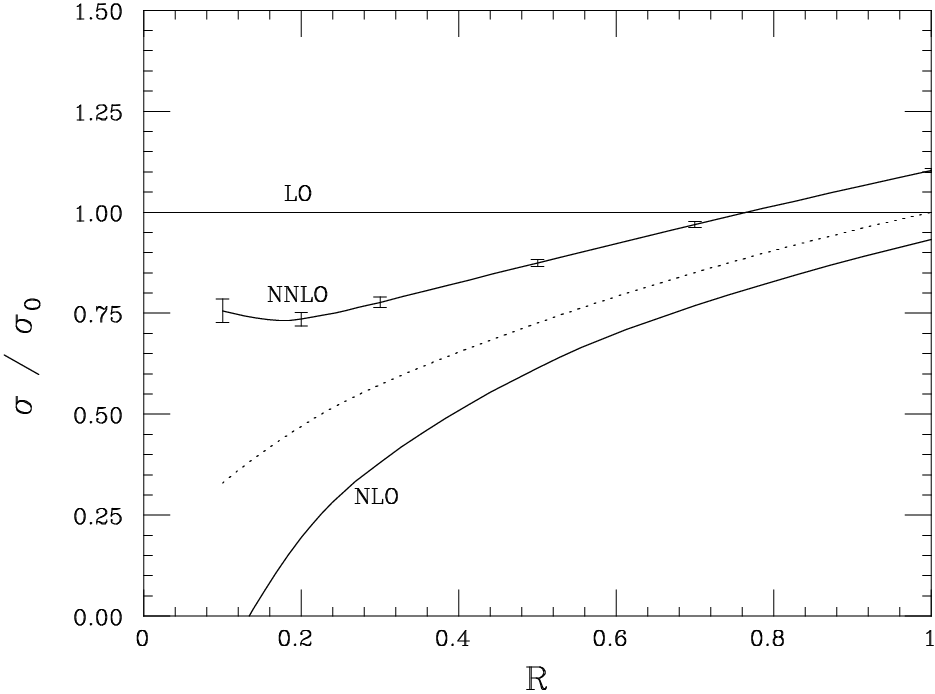}}
    }
  \caption[]{{\it The radius dependence of the inclusive jet cross
      section in the D\O\ jet algorithm with $E_0=1$~GeV in fixed-order
      (solid) and all-orders (dotted) calculations.  The error bars come
      from Monte Carlo statistics.}}
  \label{rdep}
\end{figure}

We have implemented this as a numerical integration, and find that the
cone and cluster algorithms give different results.  We concentrate on
the D\O\ algorithm, and obtain the results shown in Fig.~\ref{rdep}.
Although the absolute normalization of the corrections at each order is
outside the validity of the DLA, their dependence on physical
quantities, like the jet radius, should be well modelled.  We see that
the LO to NLO correction blows up at small $R$, and is not much better
at NNLO (where it would diverge to positive infinity if the
finite cell size, $\delta_0=0.1$, did not provide a cutoff).  This is
due to large logarithms of $R$, which arise at each order of
perturbation theory, and would need to be summed to all orders to give
accurate predictions.  This is what is done in the dotted line, which is
discussed in more detail in the next section.

\begin{figure}
  \centerline{
    \resizebox{!}{6cm}{\includegraphics{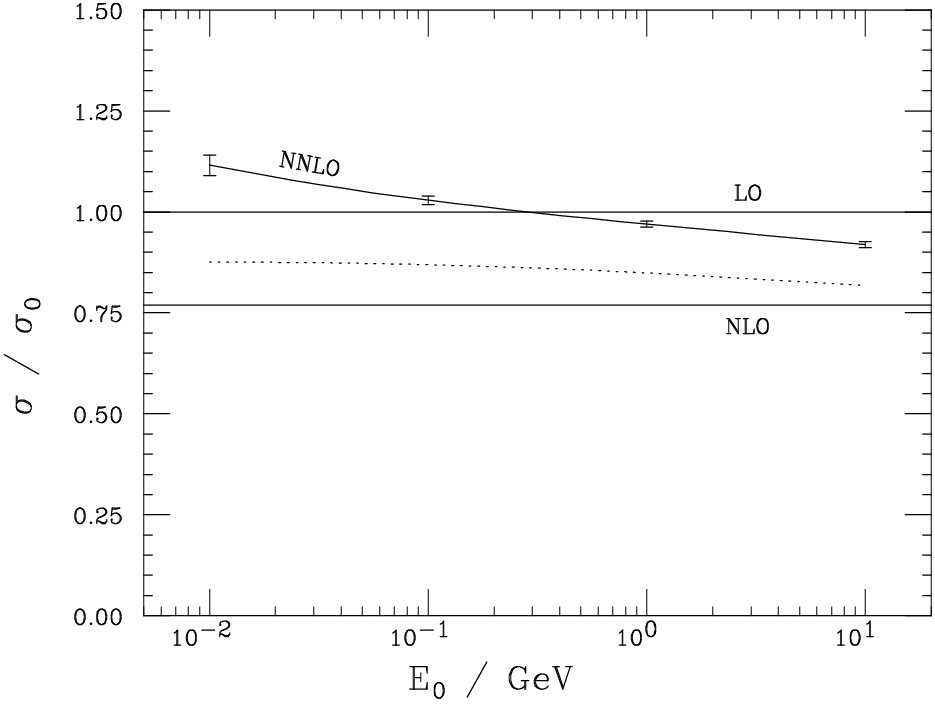}}
    }
  \caption[]{{\it The seed cell threshold dependence of the inclusive
      jet cross section in the D\O\ jet algorithm with $R=0.7$ in
      fixed-order (solid) and all-orders (dotted) calculations.}}
  \label{Edep}
\end{figure}

In Fig.~\ref{Edep} we show the $E_0$ dependence of the cross section.
It is clear that the dependence does not disappear as $E_0\to0$ at NNLO.
This,
then, is the proof that the D\O\ jet algorithm is `almost unsafe', the
safeness being rendered by the finite energy threshold of the
calorimeter.  Clearly in the ideal world of a perfect calorimeter, with
zero threshold, we get an infinite cross section.

\begin{figure}
\centerline{
  \begin{picture}(252,144)(0,0)
    \SetWidth{1.5}
    \LongArrow(126,0)(72,108)
    \LongArrow(126,0)(180,108)
    \Oval(72,108)(36,72)(0)
    \Oval(180,108)(36,72)(0)
    \Text(63,21)[c]{\large(a)}
  \end{picture}
~\hfill~
  \begin{picture}(252,144)(0,0)
    \SetWidth{1.5}
    \LongArrow(126,0)(72,108)
    \LongArrow(126,0)(180,108)
    \Oval(72,108)(36,72)(0)
    \Oval(180,108)(36,72)(0)
    \SetWidth{0.5}
    \Photon(126,36)(126,104){3}{4}
    \LongArrow(126,104)(126,108)
    \SetWidth{0.5}
    \Oval(126,108)(36,72)(0)
    \Text(63,21)[c]{\large(b)}
  \end{picture}
}
  \caption[]{{\it Illustration of the problem region for the iterative cone
      algorithm.  In (a), there are two hard partons, with overlapping
      cones.  In (b) there is an additional soft parton in the overlap
      region.}}
  \label{ill}
\end{figure}
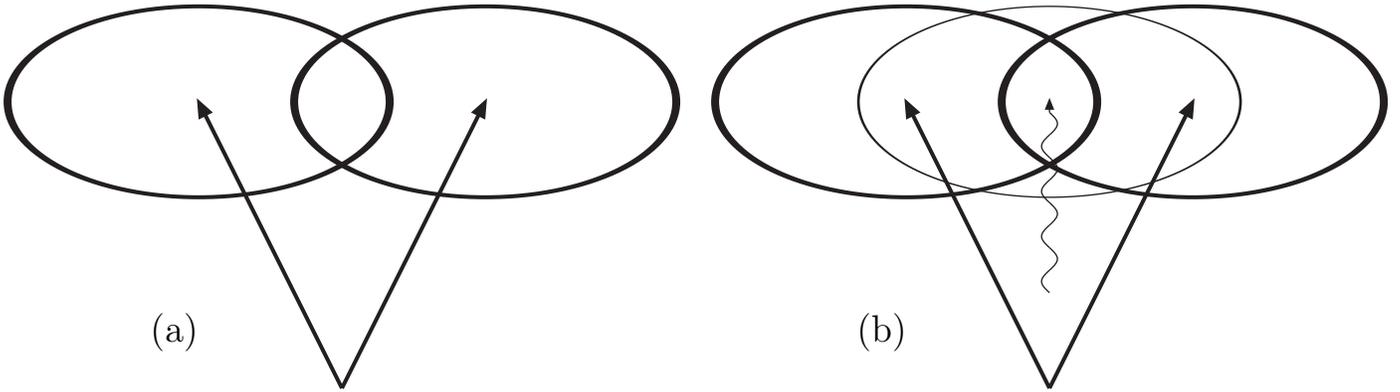

The strong $E_0$ dependence can be easily understood.  It arises from
configurations that have long been understood to be a problem in cone
algorithms, where two partons lie somewhere between $R$ and $2R$ apart
in angle, but sufficiently balanced in energy that they are both within
$R$ of their common centre, defined by Eq.~(\ref{snowmass}).  This is
illustrated in Fig.~\ref{ill}a.
According to the iterative cone algorithm, each is a separate jet,
because the cone around each seed cell contains no other active cells,
so is immediately stable.  Although the two cones overlap, there is no
energy in the overlap region, so the splitting procedure is trivial,
and it is classed as a two-jet configuration.

Now consider almost the same event, but with the addition of a soft
parton, close to the energy threshold $E_0$, illustrated in
Fig.~\ref{ill}b.  If it is marginally below threshold, the
classification is as above, with the soft parton being merged with
whichever hard parton it is nearest.  If on the other hand it is
marginally above threshold, there is an additional seed cell.  The cone
around this seed encloses both the hard partons and thus a third stable
solution is reached.  Now the merging and splitting procedure produces
completely different results.  In either the CDF or D\O\ variants the
result is the same: each of the outer jets overlaps with the central
one, with the overlap region containing 100\% of the outer one's energy.
Thus each is merged with the central one, and it is classified as a
one-jet configuration.

The classification is different depending on whether or not there is a
parton in the overlap region with energy above $E_0$.  Since the
probability for this to occur is $\sim\frac{2C_A}{\pi}\as\log E_T/E_0$,
the inclusive jet cross section depends logarithmically on the energy
threshold above which calorimeter cells are considered seed cells.  Thus
{\it the D\O\ jet definition is not fully infrared safe}.

In fact, having seen this problem we can see that it will become even
more severe at the next higher order.  Consider the probability that the
soft parton splits into two fairly collinear partons.  If they are
greater than $\delta_0$ apart, then they will fall into separate
calorimeter cells, and neither will be above $E_0$, and so a one-jet
configuration would be flipped back to a two-jet one.  The probability
for this to occur depends logarithmically on $\delta_0$, so the
inclusive jet cross section will too (at NNNLO and beyond).  Thus {\it
  the D\O\ jet definition is not fully collinear safe}.

We should restress at this point that there is nothing special about the
D\O\ algorithm in this regard.  This is a general problem of the
iterative cone type of algorithm, when used with parton directions or
calorimeter cells as seed cells.  We are picking on D\O\ simply because
they were kind enough to furnish us with full details of their
algorithm.

Before moving on, it is also worth recalling how the $\kt$ algorithm
completely avoids this issue, and remains infrared safe to all orders.
Merging starts with the softest (lowest relative $\kt$) partons.  Thus
in the configuration of Fig.~\ref{ill}b, the soft parton is first merged
with whichever hard parton it is nearer.  Only then is any decision made
about whether to merge the two jets, based solely on their opening
angle.  The algorithm has completely `forgotten about' the soft parton,
and treats the configurations of Figs.~\ref{ill}a and~\ref{ill}b
identically.  Thus, details of the calorimeter's energy threshold become
irrelevant, provided it is significantly smaller than the jet's energy.

\subsection{All-orders cross section}
\label{secall}

In the previous section, we calculated the fully inclusive jet cross
section in the DLA up to NNLO.  In fact it is possible to sum the
resulting logarithms to all orders numerically.  This is done by
constructing a parton shower algorithm, which iteratively applies the
structure seen in the first few orders of the previous section.  We have
implemented such an algorithm, incorporating exactly the same conditions
as in the fixed order calculations.  The results were shown in
Figs.~\ref{rdep} and~\ref{Edep}.  As anticipated, the poor behaviour in
the $R$-dependence at small $R$ is tamed by the resummation.
Surprisingly, the same is true of the $E_0$-dependence, which is much
milder in the all-orders result than in the NNLO result.

This can be understood as a Sudakov-type effect.  Although the fraction
of events with a hard emission in the problem region is small, the
probability of subsequent soft emission into the overlap of the cones in
those events is large, $\sim\frac{2C_A}{\pi}\as\log E_T/E_0\sim1$.  This
is precisely the logarithmic behaviour seen in the NNLO result of
Fig.~\ref{Edep}.  However, when going to the all-orders result, the
probability of non-emission exponentiates, and we obtain
\begin{equation}
  \label{suda}
  \frac{2C_A}{\pi}\as\log E_T/E_0 \longrightarrow
  1-\exp\left(-\frac{2C_A}{\pi}\as\log E_T/E_0\right)
  = 1-\left(\frac{E_0}{E_T}\right)^{\frac{2C_A}{\pi}\as},
\end{equation}
the much slower behaviour seen in the all-orders result of
Fig.~\ref{Edep}.

This result has a simple physical interpretation: in the `all-orders
environment', there are so many gluons around that there is almost
always at least one seed cell in the overlap region and the two jets are
merged to one.  Recall that in our simple approximation, the coupling
does not run.  If we retained the running coupling, this statement would
become even stronger, because the probability to emit soft gluons would
be even more enhanced.

It is precisely this effect that has lead to the belief that the merging
issue is a relatively unimportant numerical effect: \emph{in the
  experimental environment it is}.  However, expanding out the
exponential of Eq.~(\ref{suda}) as an order-by-order expansion in $\as$,
we obtain large coefficients at every order, and no hope of well-behaved
theoretical predictions.

Thus, \emph{if we are to study the internal properties of jets
  quantitatively, we must solve the overlap problem, to define jets in a
  perturbatively-calculable way.}

\subsection{The \boldmath $\Rsep$ `solution'}

It has long been realized that there is a problem with the iterative
cone algorithm due to overlapping cones.  A `solution' was
proposed[\ref{EKS3}] as follows: We know that in real life jets that
have somewhat overlapping cones will get merged, owing to the fact that
there will always be a seed cell in their overlap region.  So let's
\emph{pretend} that we are using a jet definition in which they are
merged even if there were no seed cells.  In other words, let's modify
the jet definition used in the calculation, but not in the measurement.

This modification is made by introducing an adjustable parameter, called
$\Rsep$, whereby if two partons are within an angle $\Rsep R$ of each
other, they are merged into one jet (provided that they are sufficiently
well balanced in energy to fit into the cone of radius $R$ around their
common axis).  Considerable success was claimed for this `solution' when
the same value of $\Rsep$ gave a good fit to data on both the jet cross
section as a function of cone size, and the jet shape[\ref{EKS3}]
(i.e.~the internal structure of the jet, discussed in the next section).
With the advent of more plentiful, more precise, data, it has become
necessary to make $\Rsep$ a function of the event kinematics in order to
fit data[\ref{KK}].

In our opinion, this is no solution at all.  It has a multitude of
problems:
\vspace*{-\parskip}
\begin{itemize}
\item We are calculating cross sections for one observable and comparing
  them with measured cross sections for another.  It is not clear what
  we learn from this.  In this sense the `NLO calculation' should be
  considered a model rather than a calculation of a physical cross
  section.
\item The `solution' does not generalize in any obvious way to higher
  orders.  This means that we can never learn anything quantitative
  about the internal structure of the jets, since these are only
  calculable to LO.  Furthermore it means that even NLO corrections to
  higher number of jets, such as the three-jet cross section, become
  uncalculable\footnote{In this case, an additional cut on the
    separation of the jets renders the cross section
    calculable[\ref{GK}].}.
\item The fact that we have `rescued' the NLO calculation does not imply
  that it has any formal validity.  When calculating an infrared safe
  quantity, non-perturbative corrections are suppressed by inverse
  powers of the hard scale, $E_T$.  However, by calculating a quantity
  that is not infrared safe we have no way of knowing the size of
  non-perturbative contributions.  Even in the case of almost unsafe
  quantities, we can only guarantee that non-perturbative corrections
  are suppressed by inverse powers of $E_0$.  Since this is of order the
  hadronization scale, it can give rise to 100\% corrections, and in any
  case gives corrections that do not get smaller as $E_T$ gets bigger.
\end{itemize}
This last point means that the property of asymptotic scaling, normally
expected of jet cross sections, is broken.  We should not be terribly
surprised if NLO calculations do not predict the energy-dependence of
jet cross sections very well.

\subsection{A solution}
\label{Asolution}

\begin{figure}
  \centerline{
    \resizebox{!}{6cm}{\includegraphics{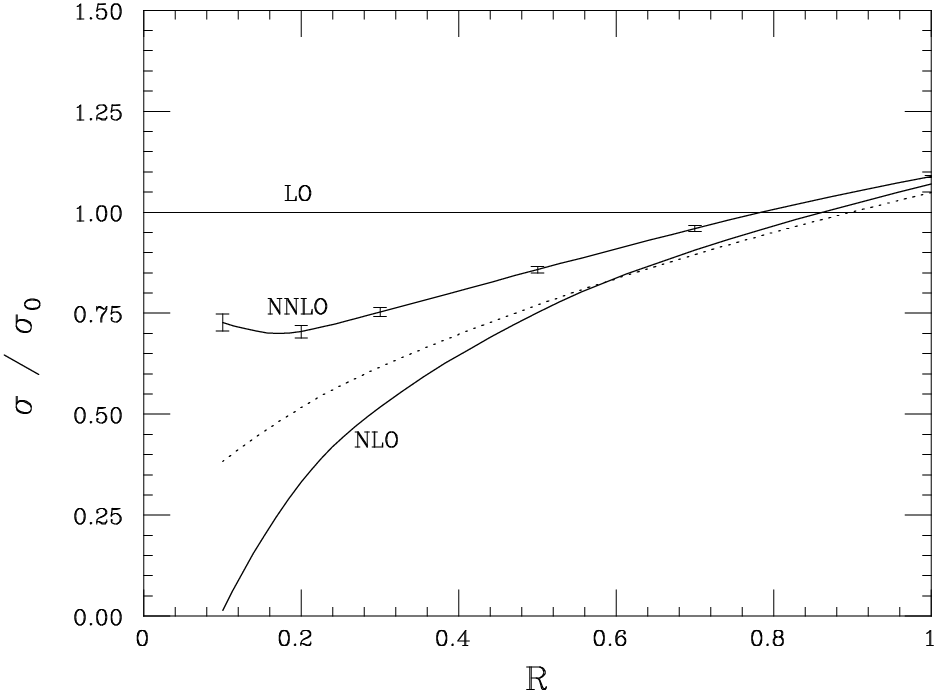}}
    \hfill
    \resizebox{!}{6cm}{\includegraphics{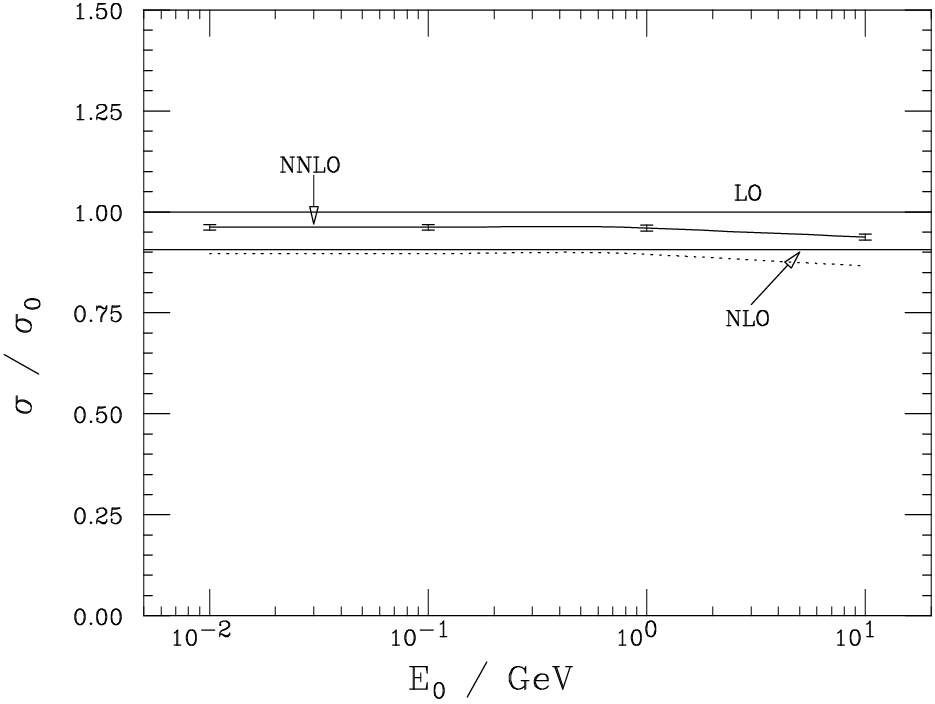}}
    }
  \caption[]{{\it The radius dependence with $E_0=1$~GeV (left) and seed
      cell threshold dependence with $R=0.7$ (right) of the inclusive
      jet cross section in the improved iterative cone algorithm, in
      which midpoints of pairs of jets are used as additional seeds for
      the jet-finding, in fixed-order (solid) and all-orders (dotted)
      calculations.}}
  \label{improved}
\end{figure}

Shortly after the $\Rsep$ `solution' was proposed, Steve Ellis
proposed[\ref{LdP}]
a solution that suffers none of the problems listed above.  It is a
simple modification to the algorithm used in both the theoretical
calculation and the experimental measurement:
\begin{quote}
  \em After finding all possible jets using the seed cells, rerun the
  algorithm using the midpoint of all pairs of jets found in the first
  stage as additional seeds\footnote{To save computer time, it is
    sufficient to just do this for jet pairs that are between $R$ and
    $2R$ apart.}.
\end{quote}
This means that the results become insensitive to whether there was a
seed cell in the overlap region, and hence to the energy threshold
$E_0$.  Cross sections are well-behaved and calculable order by order in
perturbation theory, as shown in Fig.~\ref{improved}.
Experimental results would be little changed by this modification
(compare the all-orders results of Figs.~\ref{Edep} and~\ref{improved}),
but the theoretical predictions would be enormously improved (compare
the NNLO results of Figs.~\ref{Edep} and~\ref{improved}).

It should be stressed that this does not completely remove the problem
of merging and splitting of overlapping cones.  It merely relegates it
to a procedural problem: one should state clearly the procedure one
uses, and apply it equivalently to theory and experiment.  Provided that
that procedure uses information from all the jets in a democratic way
(i.e.~not keeping track of which jets came from seed cells, and which
from the additional seeds), it will not spoil the improved properties of
the algorithm.

\subsection{A better solution}

\begin{figure}
  \centerline{
    \resizebox{!}{6cm}{\includegraphics{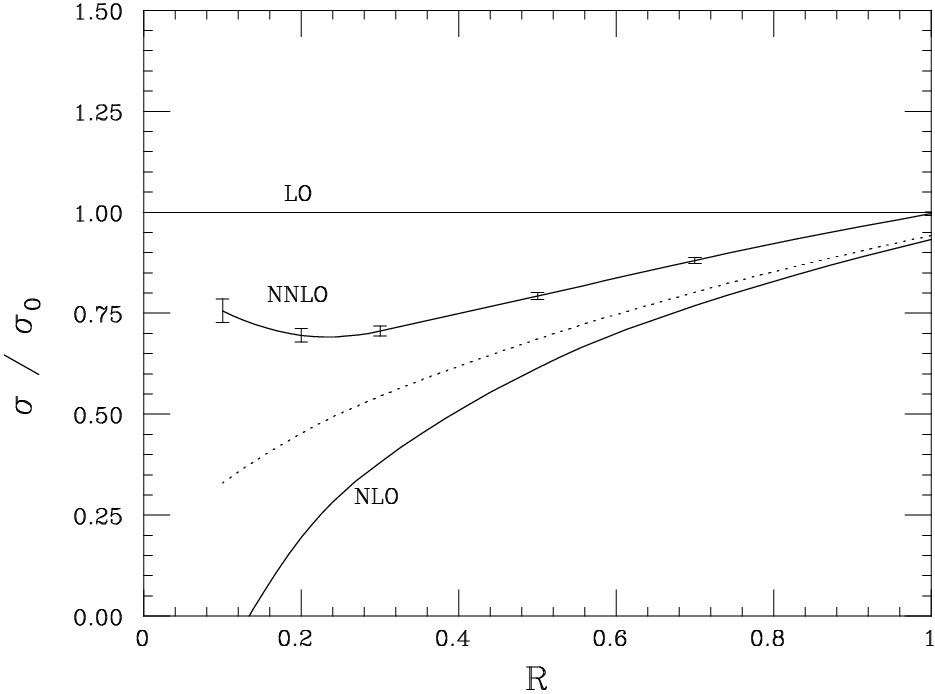}}
    }
  \caption[]{{\it The `radius' dependence of the inclusive jet cross
      section in the $\kt$ jet algorithm in fixed-order (solid) and
      all-orders (dotted) calculations.}}
  \label{kt}
\end{figure}

We finish this section by noting that using the $\kt$ algorithm removes
these problems completely.  It is fully infrared safe, and has no
overlap problem, because every final-state particle is assigned
unambiguously to one and only one jet.  We show results in
Fig.~\ref{kt}.

Unless there are factors of which we are unaware, abandoning the
iterative cone algorithm and using the $\kt$ algorithm instead would be
an even better solution than the previous one.

\section{The jet shape}

The jet shape is, at present, the most common way of resolving internal
jet structure.  It is inspired by the cone-type jet algorithm, but its
use is not restricted to cone jets.  It is defined by first running a
jet algorithm to find a jet axis.  The jet shape $\Psi(r;R)$ is then:
\begin{equation}
  \label{Psi}
  \Psi(r;R) =
  \frac{ \sum_i E_{Ti} \; \Theta(r-R_{i\mathrm{jet}}) }
       { \sum_i E_{Ti} \; \Theta(R-R_{i\mathrm{jet}}) },
\end{equation}
where the sum over $i$ can be over either all particles in the event, as
used by CDF and D\O, or only those particles assigned to the jet, as
used by ZEUS.  We have found that using cone-type jet definitions, there
is little difference between the two (less than 10\% even at the jet
edge).  However, if the jet is defined in the $\kt$ algorithm, we shall
see that there are strong reasons for preferring the definition in which
the sum is only over those particles assigned to the jet.  For now, we
concentrate on the more commonly-used definition in which the sum is
over all particles.  Thus $\Psi$ is the fraction of all energy within a
cone of size $R$ around the jet axis that is within a smaller cone of
size $r$, also around the jet axis.  Clearly we have $\Psi(R;R)=1$, with
$\Psi(r;R)$ rising monotonically~in~$r$.

In this paper, we prefer to work in terms of the differential jet shape:
\begin{equation}
  \label{psi}
  \psi(r;R) = \frac{d\Psi(r;R)}{dr}.
\end{equation}
Thus $\psi\,dr$ is the fraction of all energy within a cone of size $R$
around the jet axis that is within an annulus of radius $r$ and width
$dr$, centred on the jet axis.

To define $\psi$ in a cross sectional form, imagine defining the
single-particle production cross section, differential in the particle
$E_T$ and distance from the jet axis $r$,
\begin{equation}
  \frac{d\sigma}{dE_T\,dr}.
\end{equation}
$\psi$ is then given by:
\begin{equation}
  \label{psix}
  \psi(r;R) = \frac{\int dE_T \; E_T\frac{d\sigma}{dE_T\,dr}}
  {\int_0^R dr \int dE_T \; E_T\frac{d\sigma}{dE_T\,dr}}.
\end{equation}

\subsection{Leading order calculation}

As we have mentioned several times already, the NLO matrix elements for
the jet cross section determine the jet shape at LO.  However, it is
important to note that in taking the ratio in Eq.~(\ref{psix}), both
parts should be evaluated to the same order (LO in this case).  If the
denominator is mistakenly evaluated to NLO, one gets artificially large
renormalization scale dependence.  Compare
\begin{equation}
  \label{wrong}
  \mu^2\frac{d}{d\mu^2} \; \frac{A_0\as^3(\mu)}
  {B_0\as^2(\mu)+B_1\as^3(\mu)+2\beta_0B_0\as^3(\mu)\log\mu^2/E_T^2}
  = -3\beta_0\frac{A_0}{B_0}\as^2(\mu)+{\cal{O}}(\as^3),
\end{equation}
with the correct LO result,
\begin{equation}
  \label{right}
  \mu^2\frac{d}{d\mu^2} \; \frac{A_0\as^3(\mu)}
  {B_0\as^2(\mu)}
  = -\beta_0\frac{A_0}{B_0}\as^2(\mu)+{\cal{O}}(\as^3).
\end{equation}
This mistake was made in a recent paper on jet shapes[\ref{GGK3}], which
we discuss in more detail in a later section.

We can avoid having to use the virtual matrix elements at all, by noting
that they only contribute to $\psi(r;R)$ at exactly $r=0$.  Thus we
can calculate $\psi(r;R)$ for all $r>0$ from the tree level matrix
elements and then get the contribution at $r=0$ from the fact that it
must integrate to 1, i.e.
\begin{equation}
  \psi(r;R) = \delta(r) +
  \Bigl( \psi_{\mathrm{tree\;level}}(r;R) \Bigr)_+,
\end{equation}
where $f(r;R)_+$ is a distribution defined in terms of the function
$f(r;R)$ by $f(r;R)_+=f(r;R)$ for $r>0$ and $\int_0^Rf(r;R)_+dr=0$.
We do not therefore need a full NLO Monte Carlo program like
JETRAD[\ref{GGK1}], but instead use the tree level matrix elements from
NJETS[\ref{K}] with our own phase space generation, adapted to be
efficient for jet shapes.

It is also useful to have an analytical approximation to the matrix
elements to work with.  This can be done using the modified leading
logarithmic approximation (MLLA), in which we have contributions from
soft and/or collinear final-state emission, and soft initial-state
emission.

The probability of final-state emission from a parton of type $a$ is
given by
\begin{equation}
  dP_a = \frac12\sum_b \frac{\as}{2\pi} \, \frac{d\rho^2}{\rho^2} \,
  \frac{d\varphi}{2\pi} \, dz \, P_{a\to bc}(z),
\end{equation}
where $\varphi$ is an azimuth around the jet axis, which we always
integrate out from now on, $z$ is the fraction of $a$'s energy carried
by $b$, and $\rho$ is the opening angle between the partons.  The phase
space limits come from the requirements that both
partons be within $R$ of the jet axis, and the opening angle be less
than $\Rsep R$.  Thus we have
\begin{equation}
  \psi_a(r) = \sum_b \frac{\as}{2\pi} \, \frac2r \int_0^{1-Z} dz \,
  z\,P_{a\to bc}(z),
\end{equation}
where
\begin{equation}
  Z=\left\{ \begin{array}{ll}
      \frac r{r+R}     & \mbox{if $r<(\Rsep-1)R$,} \\
      \frac r{\Rsep R} & \mbox{if $r>(\Rsep-1)R$}
    \end{array} \right..
\end{equation}
Thus we obtain for a quark jet,
\begin{equation}
  \psi_q(r) = \frac{C_F\as}{2\pi} \, \left[ \frac2r
  \left( 2\log\smfrac1Z - \smfrac32(1-Z)^2 \right) \right]_+,
\end{equation}
and for a gluon jet
\begin{eqnarray}
  \psi_g(r) &=& \frac{C_A\as}{2\pi} \, \left[ \frac2r
  \left( 2\log\smfrac1Z - (1-Z)^2
    \left(\smfrac{11}6 - \smfrac13Z + \smfrac12Z^2\right) \right)
  \right]_+
\nonumber\\
  &+& \frac{T_RN_f\as}{2\pi} \, \left[ \frac2r
  (1-Z)^2\left(\smfrac23 - \smfrac23Z + Z^2\right) \right]_+,
\end{eqnarray}
where $N_f$ is the number of flavours.

Note that owing to the supersymmetric relations between the splitting
kernels[\ref{bible}], for $N_f=C_A$, we have
$\psi_q(r)/C_F=\psi_g(r)/C_A$.  Thus even for $N_f=5$, we expect the
shape functions to be extremely similar in shape, and only differ in
normalization.

There is also a contribution from initial-state radiation that happens
to be inside the jet cone by chance.  As shown in~[\ref{S}], this is
essentially the same for both quark and gluon jets, and is given by a
basic dipole formula (such emission can always be treated as soft), with
colour factor $C\sim C_F\sim C_A/2$.  We set $C=C_A/2$ for all numerical
results.  We have
\begin{equation}
  dP = 4\frac{C\as}{2\pi} \, d\eta \, \frac{d\phi}{2\pi} \,
  \frac{d\omega}{\omega},
\end{equation}
where $\eta$, $\phi$ and $\omega$ are the pseudorapidity, azimuth around
the beam direction, and energy of the gluon respectively.
In terms of the jet variables, this gives
\begin{equation}
  dP = 4\frac{C\as}{2\pi} \, \rho d\rho \, \frac{dz}z,
\end{equation}
and hence
\begin{equation}
  \psi_i(r) = \frac{C\as}{2\pi} \, \left[ 2 r
    \left(\frac1{Z^2}-1\right) \right]_+.
\end{equation}
Note that although there is nothing special about the jet direction, as
far as initial-state emission is concerned, this is singular as $r\to0$,
$r/Z^2\to R^2/r$.  This is due to the fact that soft partons anywhere in
the jet will pull the jet axis away from the hard parton direction.

Since $\psi_i$ is the same for both quark and gluon jets, it spoils the
nice symmetry between quark and gluon jets, since we now have
symbolically
\begin{eqnarray}
  \psi_q &=& C_F(\mathrm{FSR}) + C(\mathrm{ISR}),\\
  \psi_g &=& C_A(\mathrm{FSR}) + C(\mathrm{ISR}).
\end{eqnarray}

\begin{figure}
  \centerline{
    \resizebox{!}{6cm}{\includegraphics{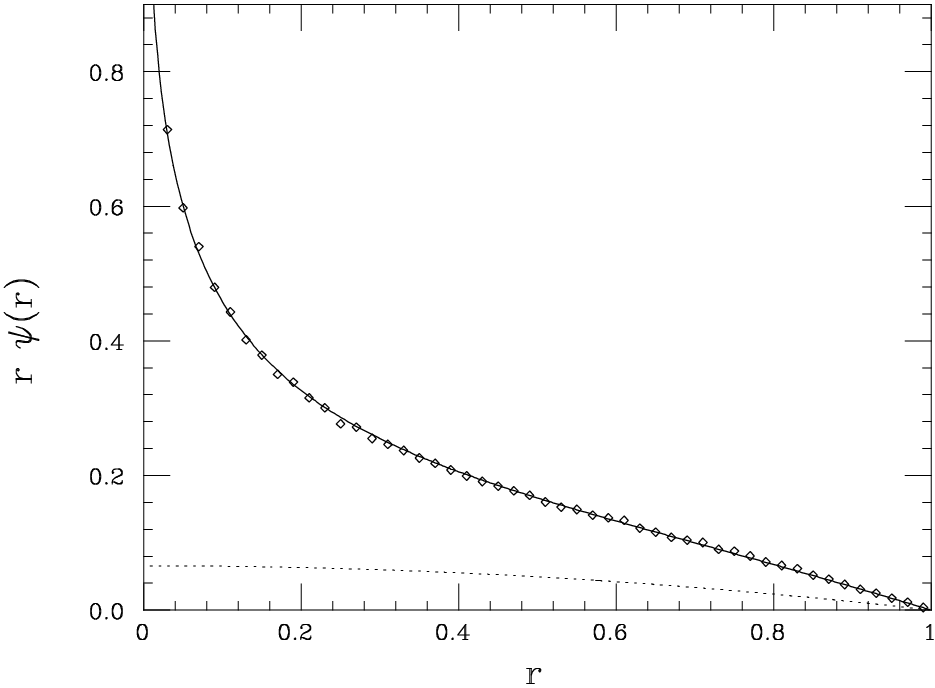}}
    \hfill
    \resizebox{!}{6cm}{\includegraphics{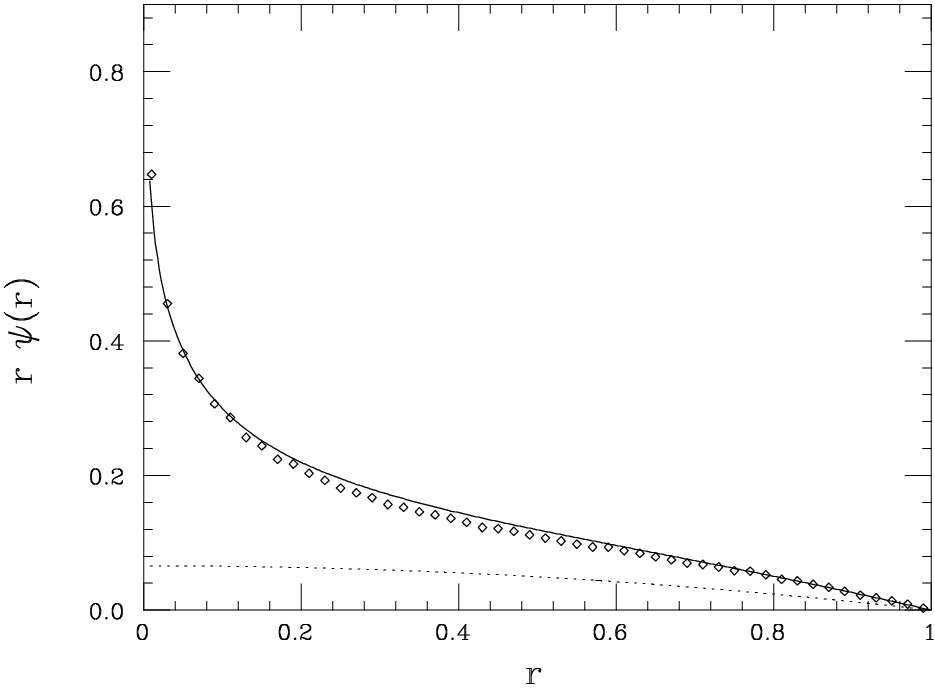}}
    }
  \caption[]{{\it The jet shape at leading order in the $\kt$ algorithm
      for a 50~GeV jet at $\eta=0$ (left) and $\eta=3$ (right),
      according to the exact tree-level matrix elements (points) and our
      analytical formulae (curves).  The contributions of the
      initial-state component of the latter are shown separately as the
      dotted curves.}}
  \label{LO}
\end{figure}

In Fig.~\ref{LO} we show the results of both the full LO matrix element
integration, and our analytical approximation to it, for the $\kt$
algorithm ($\Rsep=1$).
As with all the numerical results in this paper, we use the CTEQ4M
parton distribution functions[\ref{CTEQ}].  We see remarkably good
agreement between the full result and the analytical approximation.  The
contribution from initial-state radiation is shown separately, and is
clearly essential for this good agreement.

\begin{figure}
  \centerline{
    \resizebox{!}{6cm}{\includegraphics{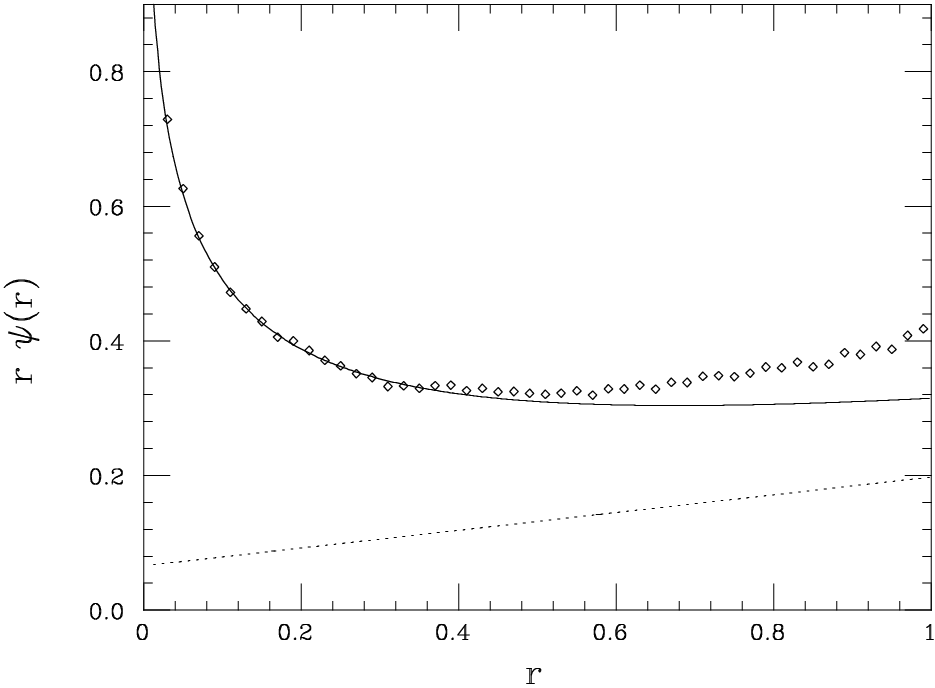}}
    \hfill
    \resizebox{!}{6cm}{\includegraphics{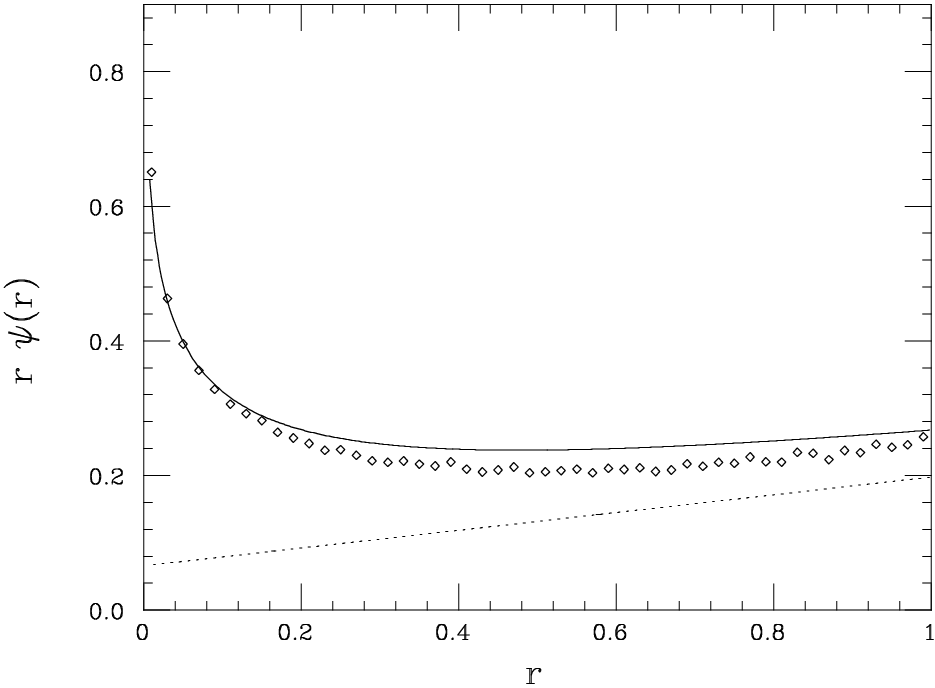}}
    }
  \caption[]{{\it As Fig.~\ref{LO}, but in the improved cone
    algorithm.}}
  \label{rsep}
\end{figure}

As $\Rsep$ increases, the approximation gets somewhat worse, as we see
in Fig.~\ref{rsep}, where we show results for the improved cone
algorithm ($\Rsep=2$).
This is not too surprising, as we are using the soft and collinear
approximations for the kinematics, and $\Rsep=2$ allows hard partons to
be up to almost $120^\circ$ apart in azimuth.  The approximation is
still pretty good though (and is still very good at $\Rsep=1.3$ for
example).

\begin{figure}
  \centerline{
    \resizebox{!}{6cm}{\includegraphics{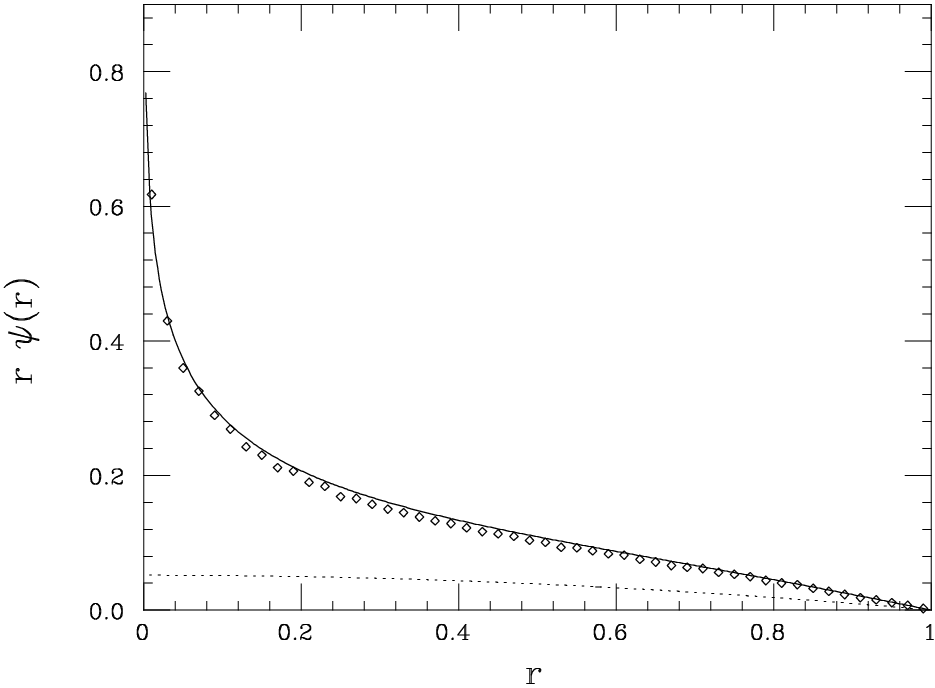}}
    \hfill
    \resizebox{!}{6cm}{\includegraphics{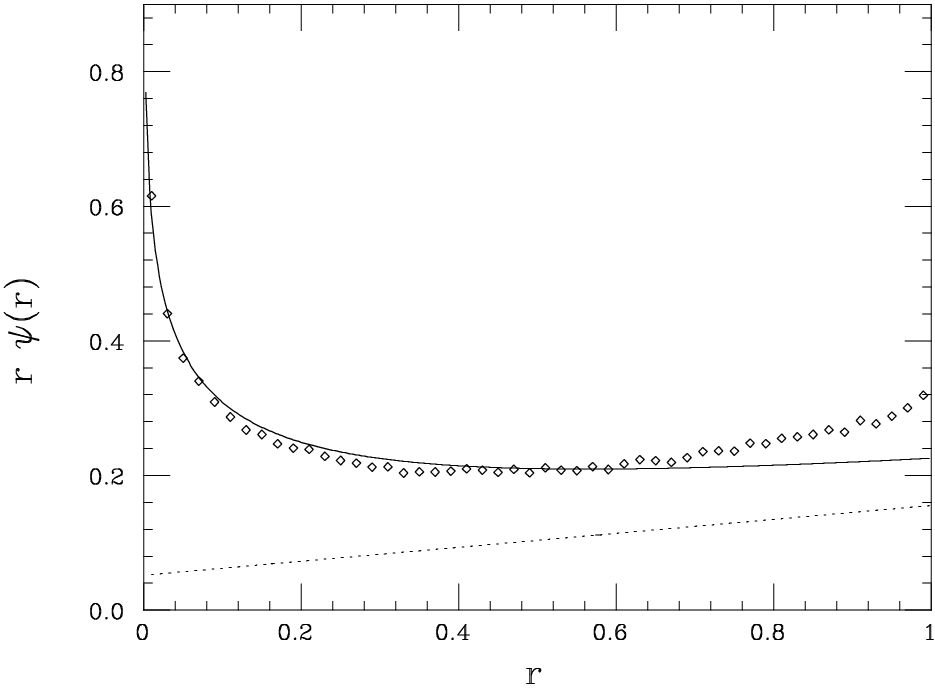}}
    }
  \caption[]{{\it The jet shape at leading order in the $\kt$ (left) and
      improved cone (right) algorithms for a 250~GeV jet at $\eta=0$.
      Points and curves are as in Fig.~\ref{LO}.}}
  \label{highET}
\end{figure}

To show that the good agreement is not just fortuitous at our chosen
$E_T$ value, we also show in Fig.~\ref{highET} the results at high
$E_T$, where the quark/gluon jet mix is very different.

\subsection{Higher orders}
\label{higher}

Having seen that the analytical results approximate the full LO matrix
element well, we move to higher orders to see how much we can improve
them.  We use the simple DLA of Sects.~\ref{secxsec} and~\ref{secall} to
estimate the effects of higher order corrections.

\begin{figure}
  \centerline{
    \resizebox{!}{6cm}{\includegraphics{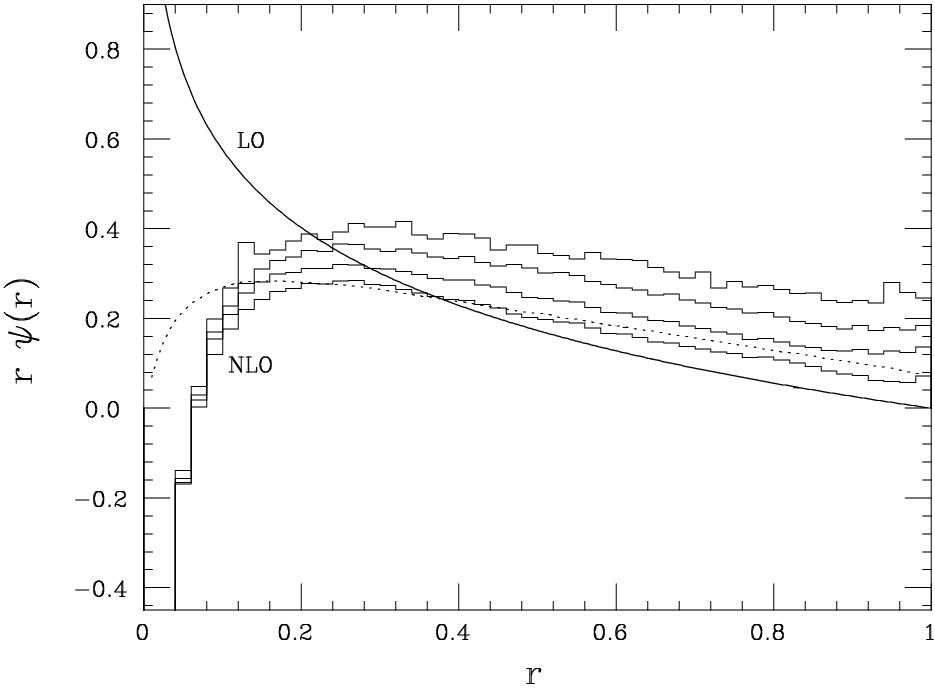}}
    }
  \caption[]{{\it The jet shape in the D\O\ jet algorithm in fixed-order
      (solid) and all-orders (dotted) calculations.  In the NLO case, we
      use $E_0=$ (from top to bottom) 0.01, 0.1, 1 and 10~GeV.}}
  \label{itshape}
\end{figure}

In Fig.~\ref{itshape} we show the jet shape in the iterative cone
algorithm.
We see that it is strongly dependent on the seed cell threshold, as
anticipated from the arguments of the previous section.  We neglect it
from further discussion.

\begin{figure}
  \centerline{
    \resizebox{!}{6cm}{\includegraphics{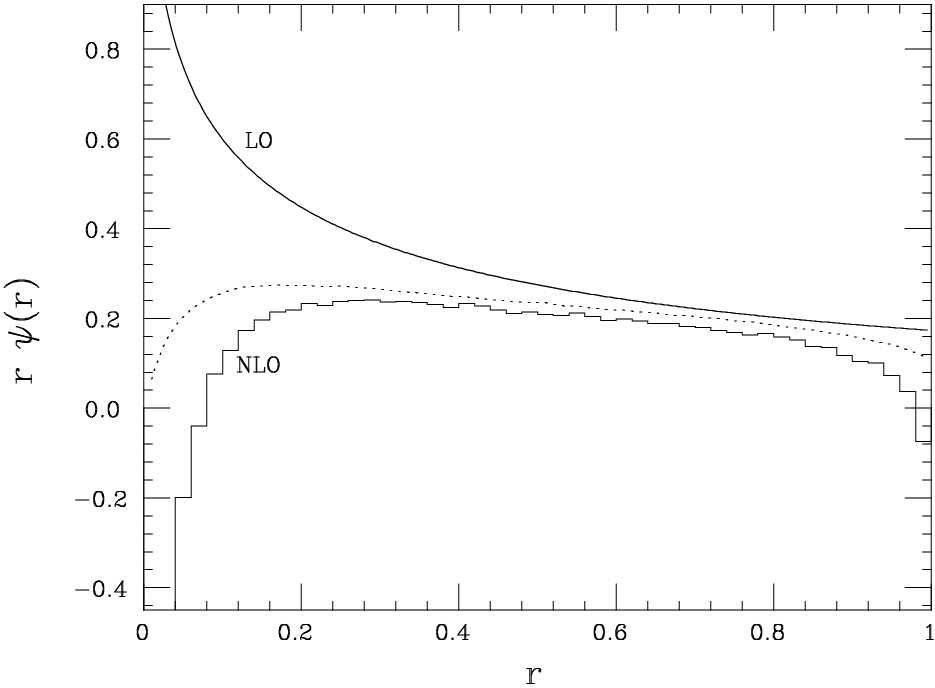}}
    \hfill
    \resizebox{!}{6cm}{\includegraphics{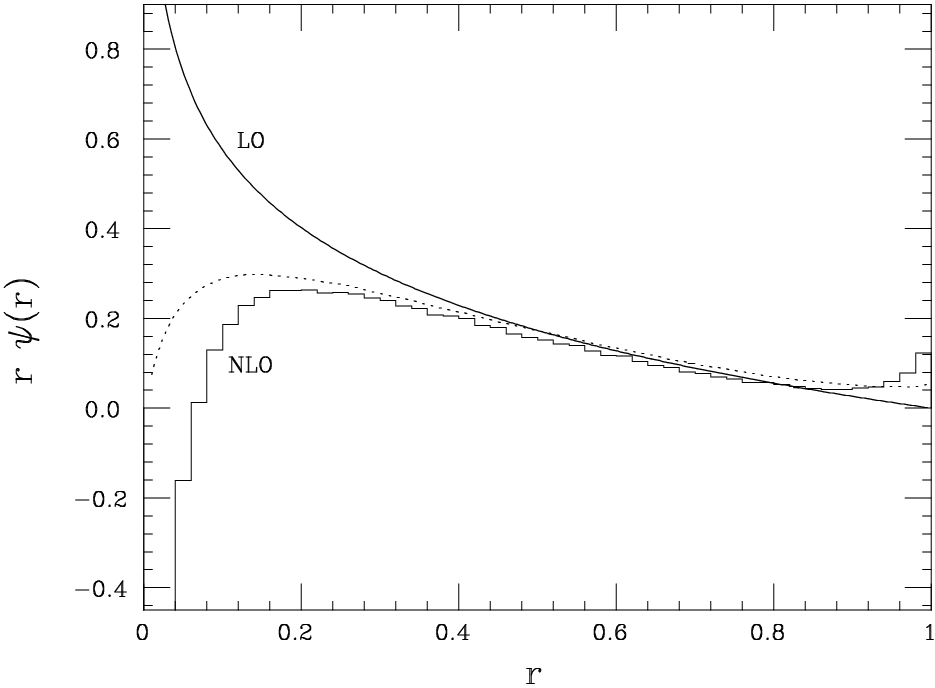}}
    }
  \caption[]{{\it The jet shape in the improved cone (left) and $\kt$
      (right) jet algorithms in fixed-order (solid) and all-orders
      (dotted) calculations.}}
  \label{ktshape}
\end{figure}

In Fig.~\ref{ktshape} we show the equivalent results in the improved
cone algorithm and the $\kt$ algorithm.
In the improved cone algorithm, the NLO corrections are rather large
(recall that the normalization is outside the control of this
approximation, and we should look at the shape of the corrections only).
Close to the jet edge, they diverge to negative infinity, a typical
`Sudakov shoulder' effect[\ref{CW}], analogous to the $C$-parameter
distribution in $e^+e^-$ annihilation for $C\sim\frac34$.  The
corresponding logarithms of $(R\!-\!r)$ must be resummed to all orders
for a reliable prediction.  The correction in both algorithms becomes
large and negative at small $r$, which we discuss in more detail in
Sect.~\ref{sudakov}.

\begin{figure}
  \centerline{
    \resizebox{!}{6cm}{\includegraphics{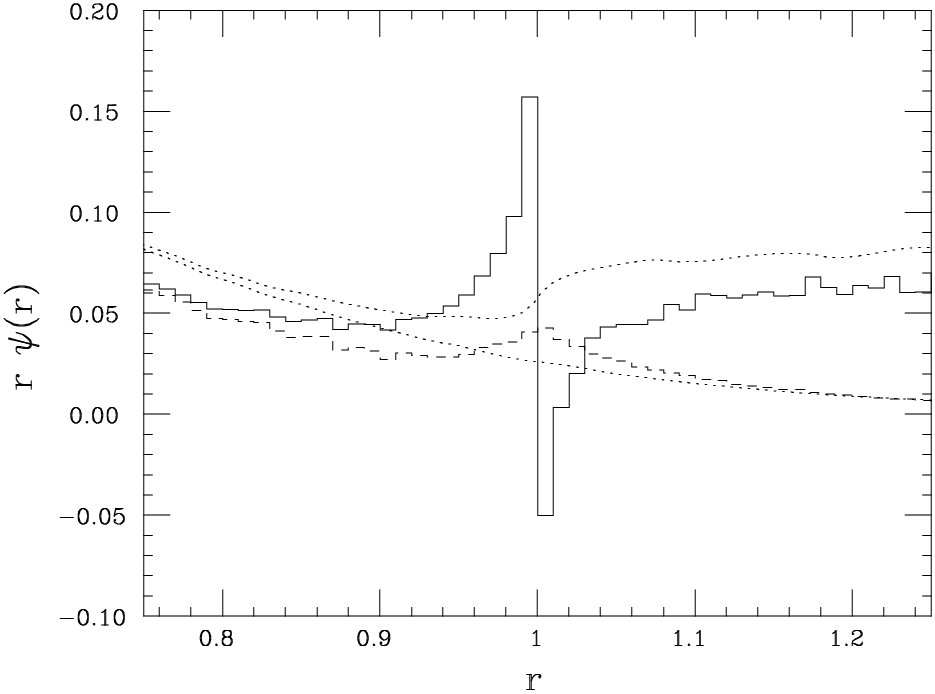}}
    }
  \caption[]{{\it The jet shape in the $\kt$ jet algorithm at NLO
      (solid and dashed) and all orders (dotted).  The solid and upper
      dotted curve use all particles in the event, the dashed and lower
      dotted curve use only those particles assigned to this jet.}}
  \label{closeup}
\end{figure}

In the $\kt$ algorithm, the NLO corrections diverge to positive infinity
near the edge of the jet.  If we take a close-up view of the region
$r\sim R$, as shown in Fig.~\ref{closeup}, we see that there is a
corresponding divergence to negative infinity just outside the jet edge.
In the region $r\sim R$, the NLO corrections are dominated by events in
which a hard gluon just outside the jet radiates a soft gluon, possibly
into the jet.  One can approximate this region analytically, and obtain
\begin{equation}
  \psi(r) \sim \left\{
    \begin{array}{lr}
      +\frac1R\log\frac{R}{R-r} & \mbox{$r<R$} \\
      -\frac1R\log\frac{R}{r-R} & \mbox{$r>R$}
    \end{array}
  \right.,
\end{equation}
exactly as seen in Fig.~\ref{closeup}.  Such terms would again have to
be summed to all orders for a reliable prediction.  We also see that
even in the all-orders calculation, $\psi(r)$ changes very rapidly with
$r$ for $r\sim R$, so this region would always be difficult to model
correctly.

In fact, this is an artifact of the fact that we define the jet shape
using all particles in the event.  If we instead use only those
particles assigned to the jet, we obtain a NLO result that is continuous
at $r\!=\!R$, although it still has an infinite change in slope there,
\begin{equation}
  \psi(r) \sim \left\{
    \begin{array}{lr}
      \frac1R-\frac{R-r}{R^2}\log\frac{R}{R-r} & \mbox{$r<R$} \\
      \frac1R-\frac{r-R}{R^2}\log\frac{R}{r-R} & \mbox{$r>R$}
    \end{array}
  \right.,
\end{equation}
again as seen in Fig.~\ref{closeup}.  This still indicates the formal
need for a resummed calculation, but the numerical result of that
resummation is a relatively minor correction, as indicated by the
difference between the all orders and NLO curves.  Note that the all
orders calculation is smooth in this case.

To avoid these large higher order terms, we recommend that in future the
jet shape be defined using \emph{only those particles assigned to the
  jet by the jet algorithm}.  Using this definition, we see that for the
$\kt$ algorithm, the NLO curve is a good approximation to the all-orders
curve for all $r\gtap0.2$.  For $r\ltap0.9$ the difference between the
two definitions is small.

\subsection{Scale of the running coupling}

As is well known, the `best' scale to use for the running coupling is
the maximum transverse momentum of emitted gluons, relative to the axis
of the emitter.  For the jet shape at angle $r$, this is different for
initial-state and final-state emission\footnote{This is why it is
  dangerous to use such schemes in exact matrix element calculations,
  since they do not separate emission into different components.}.  For
the final-state case it is $\mu = r(1-Z)E_T$, where $E_T$ is the
transverse momentum of the jet, and for the initial-state case it is
$\mu = (1-Z)E_T$.

\begin{figure}
  \centerline{
    \resizebox{!}{6cm}{\includegraphics{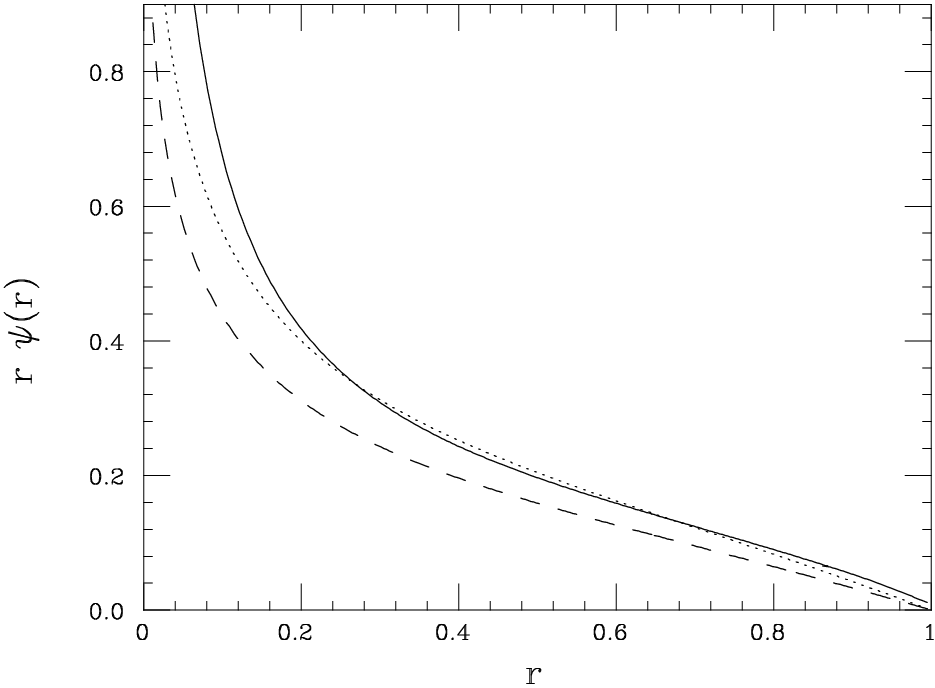}}
    \hfill
    \resizebox{!}{6cm}{\includegraphics{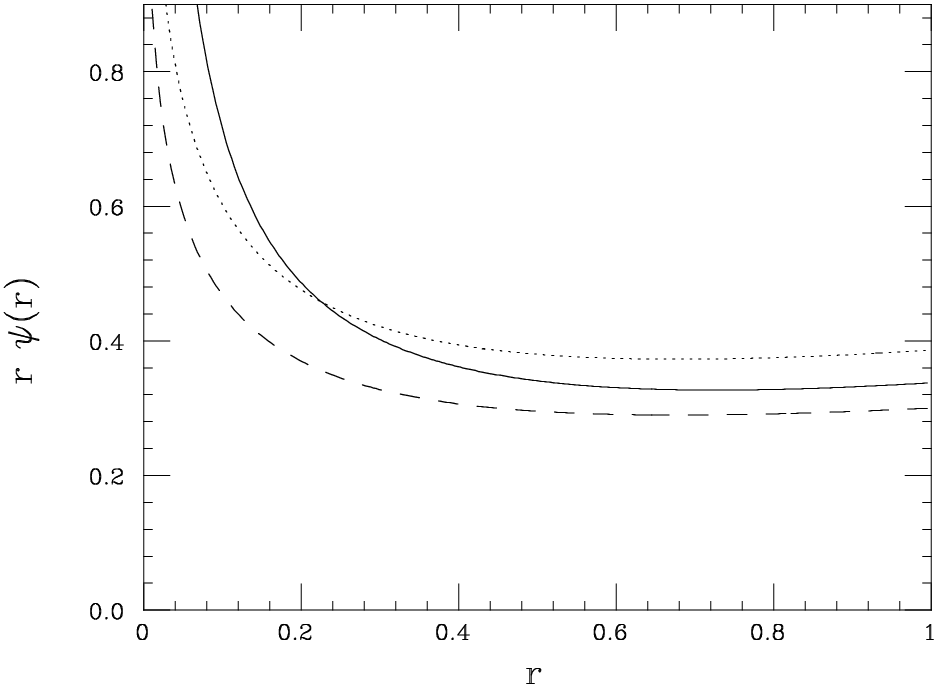}}
    }
  \caption[]{{\it Effect of running coupling on the shape of a 50~GeV
      jet in the $\kt$ (left) and improved cone (right) algorithms:
      $\mu=E_T$ (dashed), $\mu=E_T/4$ (dotted) and the `best' scale
      defined in the text (solid).}}
  \label{run}
\end{figure}

We show the results in Fig.~\ref{run}.
Relative to the default scale, $\mu=E_T$, this makes a big difference,
but much of that difference can be absorbed by choosing instead
$\mu=E_T/4$.  Then only at very small $r$ does it become important.

\subsection{Power corrections}

In incorporating running coupling effects, one can go even further, by
including $\as(q_t)$ inside the $z$ integration.  However, since the $z$
integral runs from $0$ to $1\!-\!Z$, this involves integrating over the
Landau pole, which induces power corrections.

We basically follow the philosophy of Dokshitzer and Webber[\ref{DW1}],
and begin by reviewing their method.  Consider an integral
\begin{equation}
  I_p \equiv \frac1{Q^{p+1}} \int_0^Q dk \, k^p \, \as(k).
  \label{Ip}
\end{equation}
Making a perturbative expansion, one obtains
\begin{equation}
  I_{p,\mathrm{pert}} \equiv \frac1{p+1}\left(\as
  + 2\beta_0\as^2\left(\frac1{p+1}+\log\frac\mu{Q}\right)
  + {\cal{O}}(\as^3)\right),
\end{equation}
where $\as\equiv\as(\mu)$ and $\beta_0=(11C_A-2N_f)/12\pi$.

If one introduces the notion of an infrared-finite coupling, then the
integral
\begin{equation}
  \ab(Q_0) \equiv \frac{p+1}{Q_0^{p+1}} \int_0^{Q_0} dk \, k^p \, \as(k),
\end{equation}
is a well-defined, but fundamentally non-perturbative, quantity.  Thus
we can split the integration region of Eq.~(\ref{Ip}) in two, and obtain
\begin{equation}
  I_p \equiv I_{p,\mathrm{pert}} + I_{p,\mathrm{pow}},
\end{equation}
with
\begin{equation}
  I_{p,\mathrm{pow}} = \left(\frac{Q_0}{Q}\right)^{p+1}\frac1{p+1}\left(
    \ab(Q_0) - \as - 2\beta_0\as^2\left(\frac1{p+1}+\log\frac\mu{Q_0}\right)
  + {\cal{O}}(\as^3)\right).
\end{equation}
Two values of $\abz(Q_0)$ were given in~[\ref{DW1}], fitted from the
average value of thrust in $e^+e^-$ annihilation:
\begin{eqnarray}
  \abz(\mbox{2 GeV}) &=& 0.52\pm0.03, \\
  \abz(\mbox{3 GeV}) &=& 0.42\pm0.03.
\end{eqnarray}
These numbers actually refer to $\as$ defined in the physical scheme,
rather than $\overline{\mathrm{MS}}$, but the difference is smaller than
the errors, so we ignore it.  Also, their analysis is only strictly valid
when the NLO correction is included, which it is not here.
Nevertheless, we use it to get a rough estimate of the importance of
these corrections.

In our case, for final-state emission, we have
\begin{eqnarray}
  \psi(r) &=& \frac1{2\pi} \, \frac2r \int_0^{1-Z} dz \,
  z\left(P(z)+P(1-z)\right) \as(zrE_T) \\
  &\sim& \frac{2C_i}{2\pi} \, \frac2r \, \frac1{rE_T}\int_0^{r(1-Z)E_T} dk \,
  \as(k),
\end{eqnarray}
where $C_i=C_F\,(C_A)$ for a quark (gluon) jet, and the approximation
corresponds to small $z$ (i.e.~small $k$).  Thus we have
\begin{equation}
  \psi_{\mathrm{pow}}(r) = \frac{2C_i}{2\pi} \, \frac2r \,
  \left(\frac{Q_0}{rE_T}\right)\left(
    \abz(Q_0) - \as - 2\beta_0\as^2\left(1+\log\frac\mu{Q_0}\right)
  \right).
\end{equation}
For initial-state emission, we have
\begin{equation}
  \psi_i(r) \sim \frac{2C}{2\pi} \, 2r \, \frac1{E_T}\int_0^{(1-Z)E_T} dk \,
  \as(k),
\end{equation}
and hence
\begin{equation}
  \psi_{i\mathrm{pow}}(r) = \frac{2C}{2\pi} \, 2r \,
  \left(\frac{Q_0}{E_T}\right)\left(
    \abz(Q_0) - \as - 2\beta_0\as^2\left(1+\log\frac\mu{Q_0}\right)
  \right).
\end{equation}
Note that near the core of the jet, the initial-state term is completely
insignificant, while near the edge of the jet it is comparable to the
final-state term.

\begin{figure}
  \centerline{
    \resizebox{!}{6cm}{\includegraphics{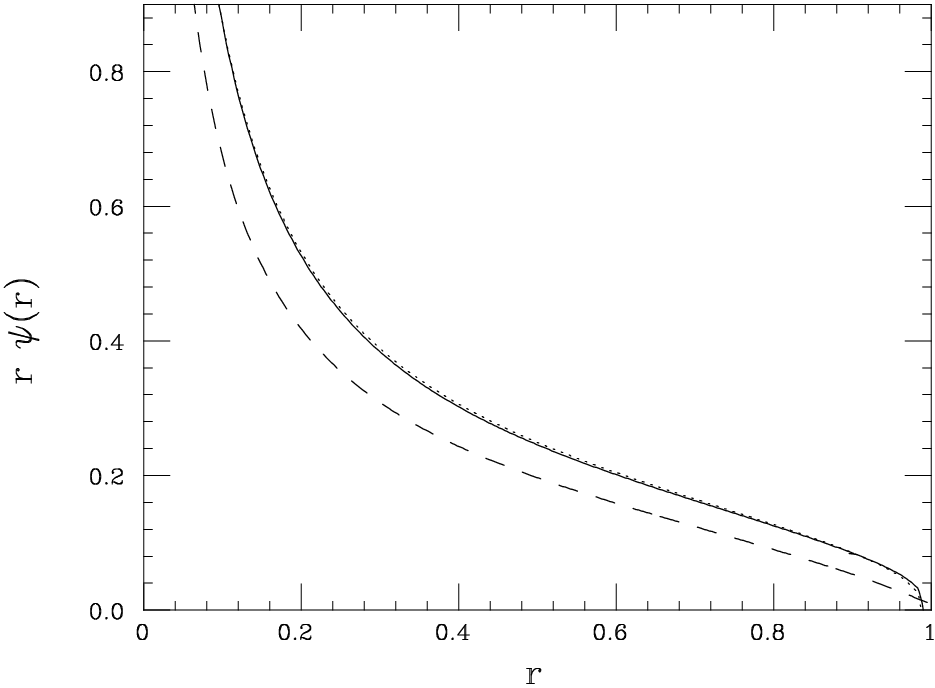}}
    \hfill
    \resizebox{!}{6cm}{\includegraphics{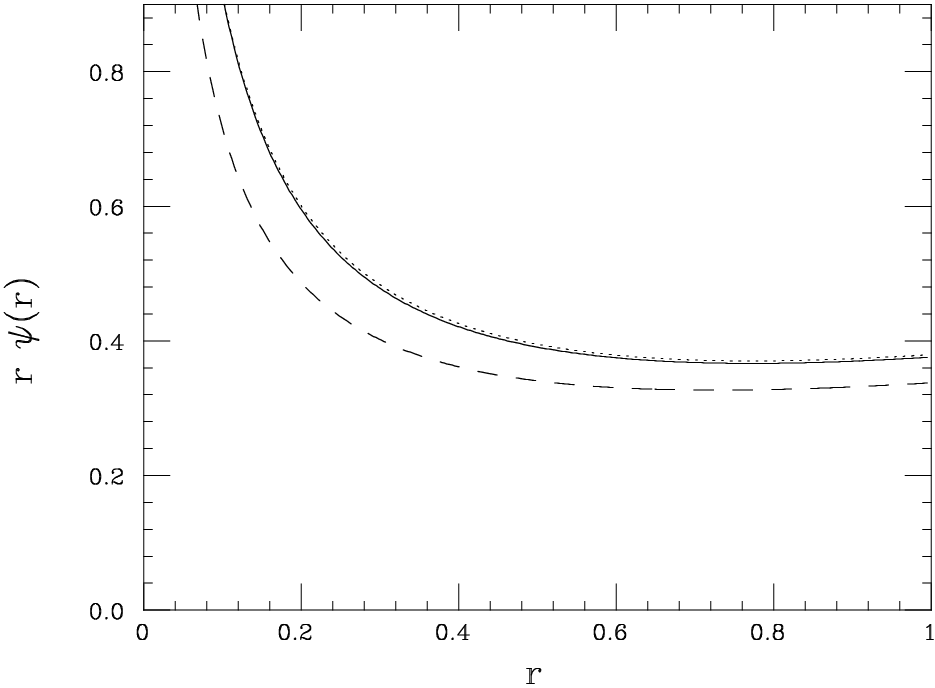}}
    }
  \caption[]{{\it Effect of power corrections on the shape of a 50~GeV
      jet in the $\kt$ (left) and improved cone (right) algorithms: no
      corrections (dashed), $Q_0=2$~GeV (solid) and $Q_0=3$~GeV
      (dotted).}}
  \label{power}
\end{figure}

Adding these non-perturbative corrections to our previous perturbative
results, we obtain the results of Fig.~\ref{power}.
We see that the power corrections are sizeable, comparable to the
anticipated size of NLO corrections.  The curves for the two values of
$\abz(Q_0)$ are completely indistinguishable~-- a cross-check of the
consistency of the calculation.

In this context, it is relevant to question what $\Rsep$ really means,
if anything.  The value usually used, $\Rsep=1.3$, was obtained by
fitting a LO calculation to the data, and perhaps it is double counting
to continue to use this value when power corrections are included.  In
fact, we find that the normalization of the power-corrected LO with
$\Rsep=1.1$ is comparable to that of LO with $\Rsep=1.3$.

\subsection{Sudakov resummation}
\label{sudakov}

In the region $r\to0$, the energy profile function develops large
logarithms at all higher orders, $\sim\as^n\log^{2n-1}r$.  These can be
resummed to at least MLLA accuracy and, we expect, to NLLA.  We work to
MLLA accuracy, i.e.~incorporating the $\sim\as^n\log^{2n-2}r$ terms as
well, but none beyond that.

As usual in these types of calculations, it is easiest to work in terms
of the total energy within $r$, $\Psi(r)$, rather than the energy
distribution at $r$, $\psi(r)$, and then differentiate afterwards.
Since to MLLA soft gluons do not carry away any momentum, it is
straightforward to see that this is actually the same as the probability
that the hard parton remains within $r$.  To leading log accuracy, this
is given by
\begin{equation}
  P(<r) = \exp\Bigl(-P_1(>r)\Bigr),
\end{equation}
where $P_1(>r)$ is the leading logarithm of the leading order in $\as$
probability that the hard parton is outside $r$.  This is improved to
MLLA accuracy simply by including the next-to-leading logarithm and
running coupling in $P_1$, giving
\begin{equation}
  P_{1q} = \int_{rE_T}^{RE_T} \frac{dq_t}{q_t}\frac{\as(q_t)}{\pi}
    \left(2\,C_F\log\frac{RE_T}{q_t}-\frac32C_F\right),
\end{equation}
for quark jets, and
\begin{equation}
  P_{1g} = \int_{rE_T}^{RE_T} \frac{dq_t}{q_t}\frac{\as(q_t)}{\pi}
    \left(2\,C_A\log\frac{RE_T}{q_t}-\frac12b_0\right),
\end{equation}
for gluon jets, where $b_0=4\pi\beta_0=\frac{11}3C_A-\frac43T_RN_f$.
Note that to this accuracy, all jet definitions are the same, and
independent of $\Rsep$.  This is because we can take the small $r$ phase
space limit, $Z\sim r/R$.

We obtain\footnote{Note that these results are identical to those
  of~[\ref{T}] for the energy-energy correlation in the back-to-back
  region of $e^+e^-$ annihilation, evaluated to the same accuracy
  (i.e.~their Eq.~(10) plus the second line of Eq.~(11)), except that
  theirs is a factor of 2 larger, because they have two jets while we
  have~one.}
\begin{eqnarray}
  P_q(r) &=& \exp\left\{2C_F\log\frac{R}{r}
    \;f_1\left(2\as\beta_0\log\frac{R}{r}\right)
    -\smfrac32C_F \;f_2\left(2\as\beta_0\log\frac{R}{r}\right)
  \right\}, \\
  P_g(r) &=&\exp\left\{2C_A\log\frac{R}{r}
    \;f_1\left(2\as\beta_0\log\frac{R}{r}\right)
    -\smfrac12b_0 \;f_2\left(2\as\beta_0\log\frac{R}{r}\right)
  \right\},
\end{eqnarray}
where $\as=\as(RE_T)$, and
\begin{eqnarray}
  f_1(\lambda) &=& \frac1{2\pi\beta_0\lambda}
  \left(\log(1-\lambda)+\lambda\right), \\
  f_2(\lambda) &=& \frac1{2\pi\beta_0}
  \log(1-\lambda).
\end{eqnarray}
Differentiating with respect to $r$, we obtain
\begin{eqnarray}
  \psi_q &=& \frac{\as(rE_T)}{2\pi} \, \frac2r \,
  \left(2\,C_F\log\frac{R}{r}-\frac32C_F\right) P_q, \\
  \psi_g &=& \frac{\as(rE_T)}{2\pi} \, \frac2r \,
  \left(2\,C_A\log\frac{R}{r}-\frac12b_0\right) P_g.
\end{eqnarray}
Noting that $P_q=P_g+{\cal{O}}(\as)=1+{\cal{O}}(\as)$, it is
straightforward to see that these agree with the small-$r$ limits of the
functions derived earlier.

The initial-state radiation that happens to be within the jet also
contributes to MLLA accuracy, so must also be included.  To fully
include this to NLLA accuracy would require a more complicated analysis
since, as noted earlier, the ISR and FSR contributions have different
coupling effects.  To MLLA however, it can be done in a straightforward
way, simply by modifying $\frac32C_F \to \frac32C_F-C\,R^2$ and
$\frac12b_0 \to \frac12b_0-C\,R^2$.

We match these functions with their leading order counterparts whenever
we use them (i.e.~we subtract off their leading order expansions in
$\as$).  In principle this could induce unphysical behaviour at large
$r$, since these resummed formulae are not valid there, but in fact we
find that they tend to zero like $R\!-\!r$ at $r\to R$, where the
leading order result is either non-zero, for $\Rsep>1$, or has the same
behaviour, for $\Rsep=1$.  Thus we do not have to take any special
precautions at the edge of phase space.

\begin{figure}
  \centerline{
    \resizebox{!}{6cm}{\includegraphics{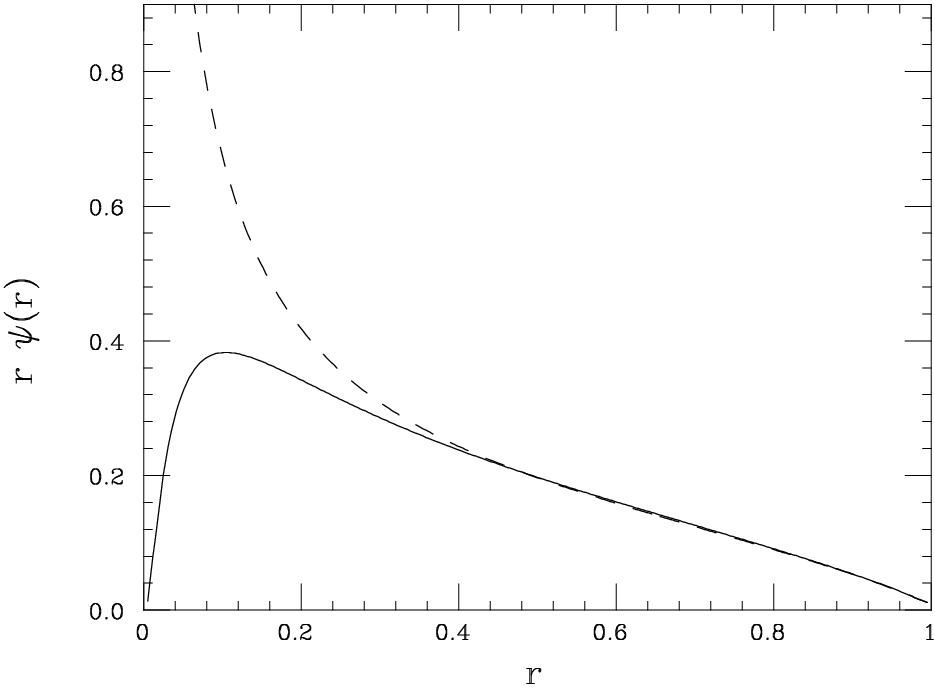}}
    \hfill
    \resizebox{!}{6cm}{\includegraphics{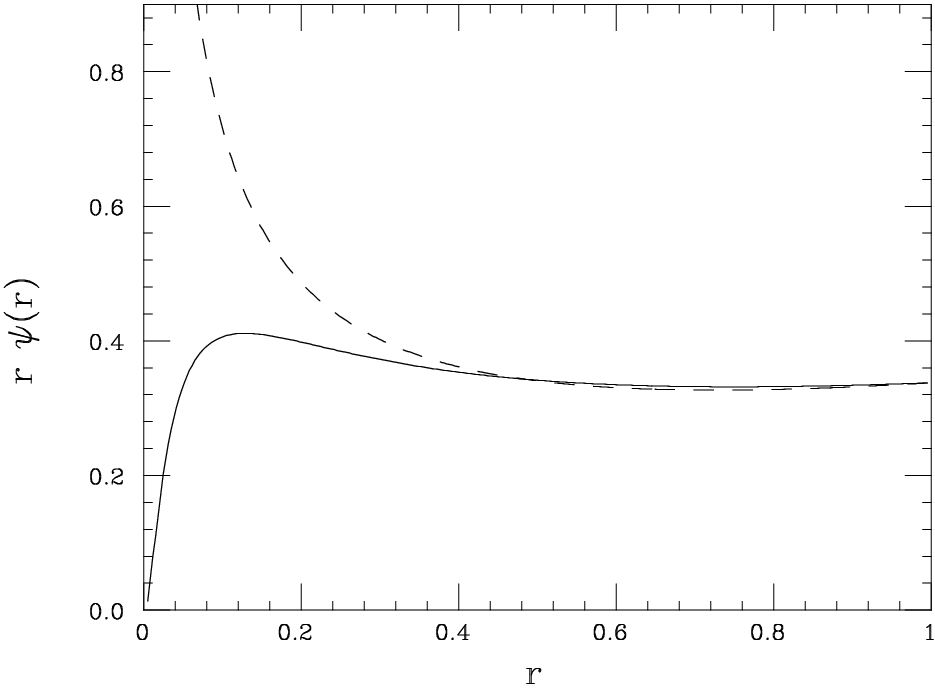}}
    }
  \caption[]{{\it Effect of resummation on the shape of a 50~GeV
      jet in the $\kt$ (left) and improved cone (right) algorithms: with
      (solid) and without (dashed) resummation of logarithms of $R/r$.}}
  \label{resum}
\end{figure}

We show the results in Fig.~\ref{resum}, where the resummation is seen
to be crucial at small $r$, taming the unphysical rise in $\psi$, but it
does not make a great deal of difference for $r>0.2$, where most of the
data are.
Note that our result should not be trusted for very small $r$, i.e.~for
$\lambda\sim1$, since higher order effects such as the emission of a
pair of gluons with balancing transverse momentum must then become
important[\ref{RW}].  This is signalled by the divergence of $f_1$ and
$f_2$ at $\lambda=1$.

One might expect that the integral inside the form factor receives power
corrections, as has been observed for some event shapes in $e^+e^-$
annihilation[\ref{DW2}].  However, following the method of~[\ref{DW2}],
one finds only sources of $1/E_T^2$ corrections in our case, which are
insignificant relative to the $1/E_T$ terms already discussed.  Thus we
do not bother including them.  We do however assume that the leading
$1/E_T$ corrections from final-state radiation are suppressed by the
perturbative form factor, since soft gluon emission should only be
enhanced at small angles in events in which the jet initiator has
remained at a small angle itself.

\begin{figure}
  \centerline{
    \resizebox{!}{6cm}{\includegraphics{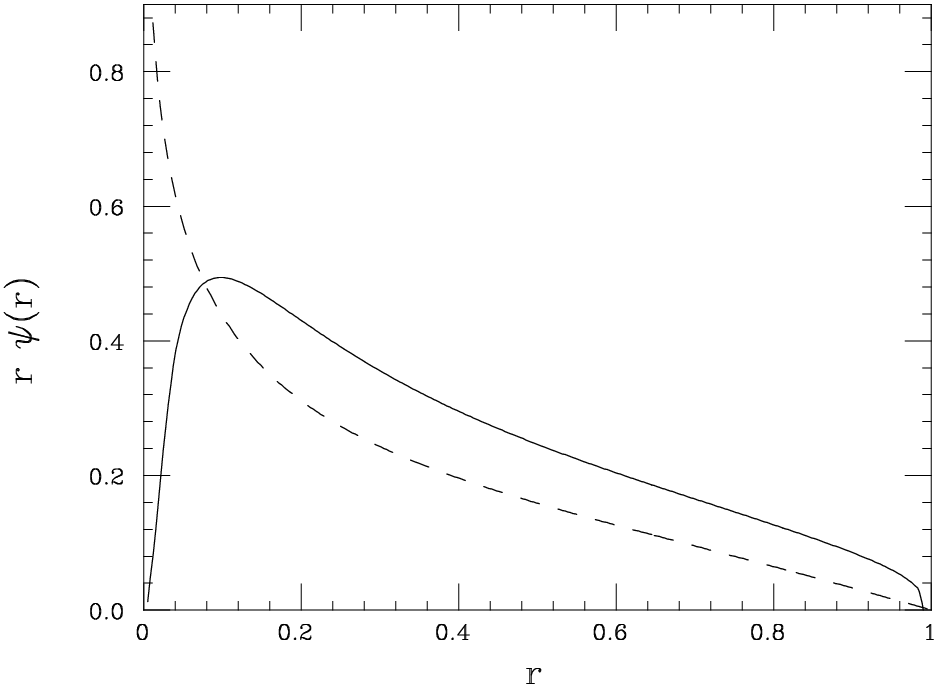}}
    \hfill
    \resizebox{!}{6cm}{\includegraphics{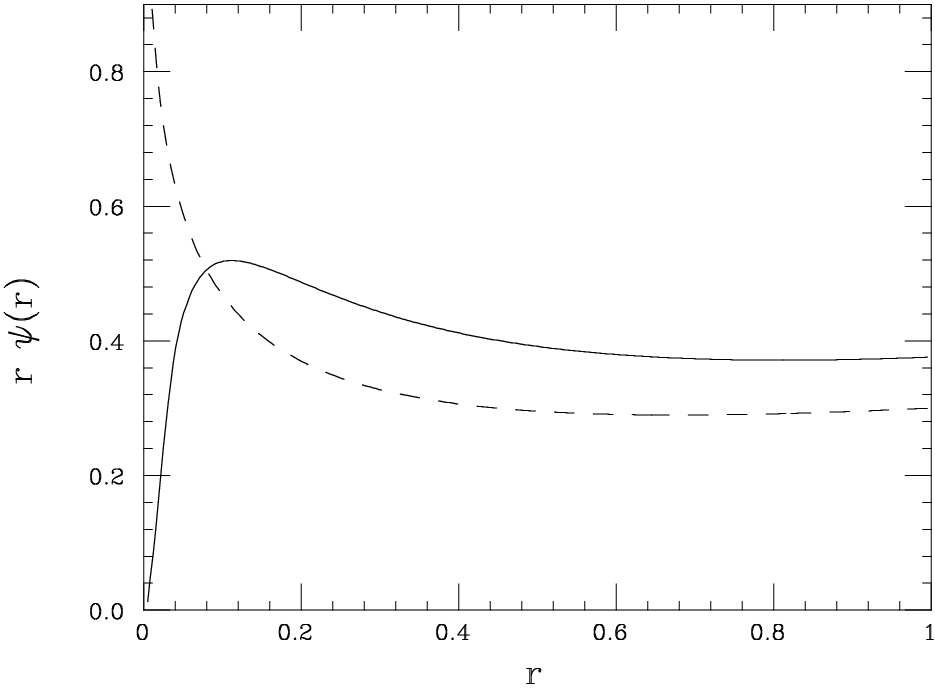}}
    }
  \caption[]{{\it Total effect of running coupling, power corrections
      and resummation on the shape of a 50~GeV jet in the $\kt$ (left)
      and improved cone (right) algorithms: LO (dashed) and with
      everything (solid).}}
  \label{final50}
\end{figure}
\begin{figure}
  \centerline{
    \resizebox{!}{6cm}{\includegraphics{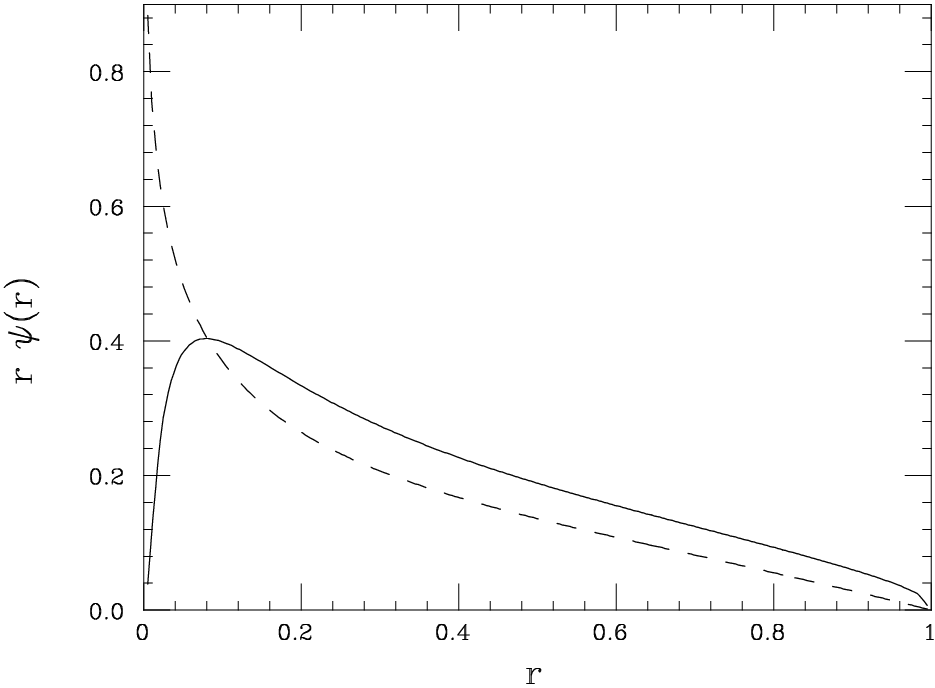}}
    \hfill
    \resizebox{!}{6cm}{\includegraphics{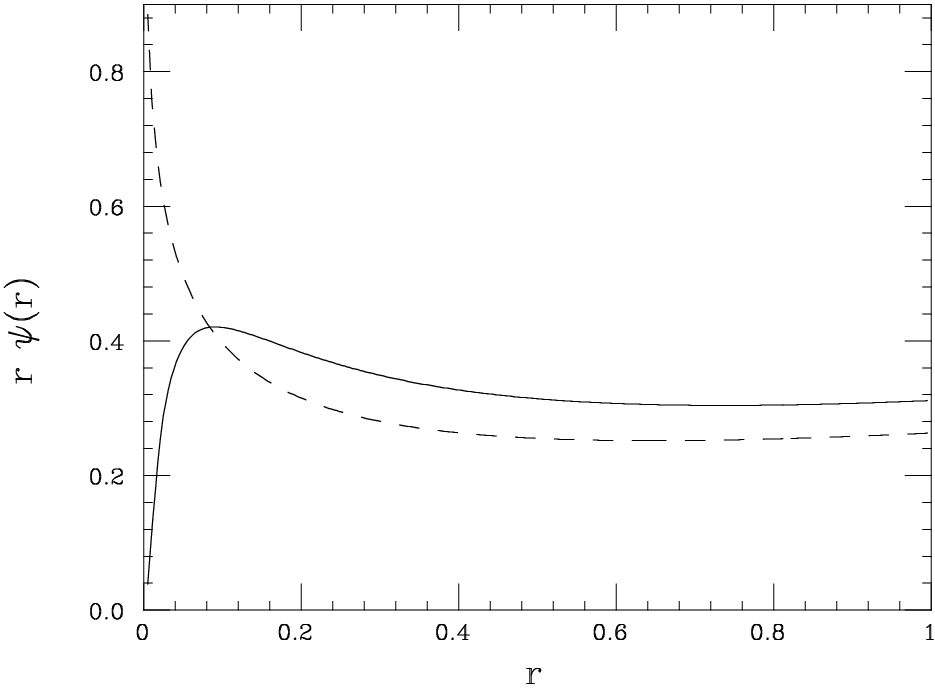}}
    }
  \caption[]{{\it As Fig.~\ref{final50} but for a 100~GeV jet.}}
  \label{final100}
\end{figure}
\begin{figure}
  \centerline{
    \resizebox{!}{6cm}{\includegraphics{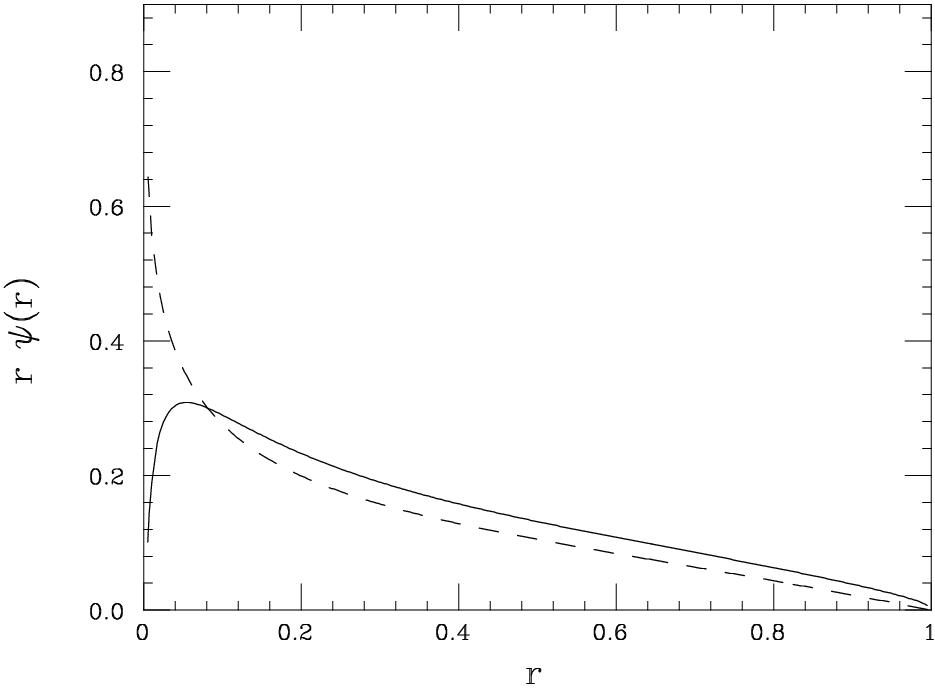}}
    \hfill
    \resizebox{!}{6cm}{\includegraphics{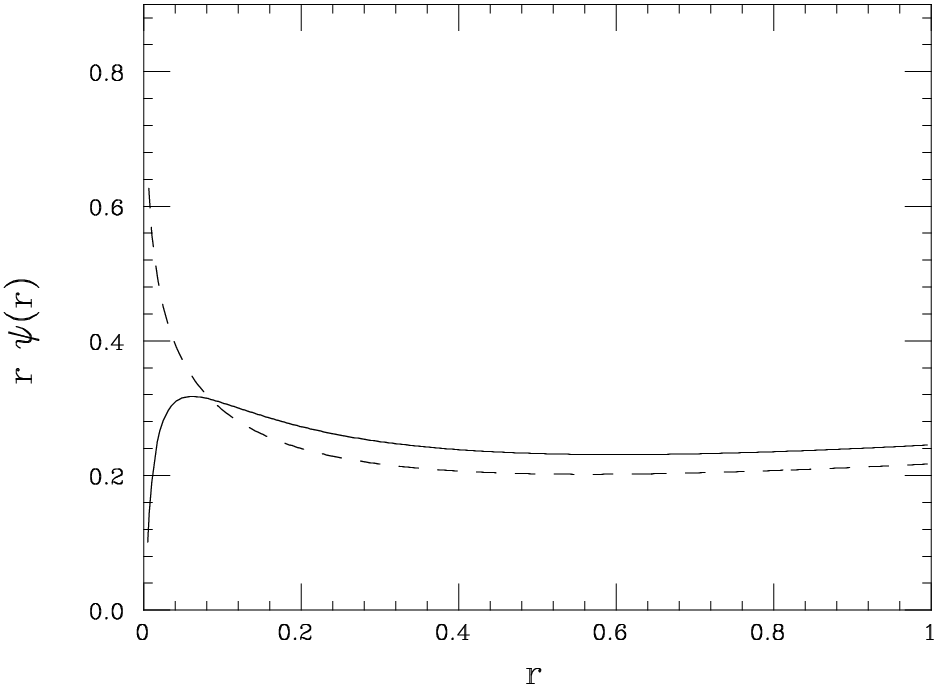}}
    }
  \caption[]{{\it As Fig.~\ref{final50} but for a 250~GeV jet.}}
  \label{final250}
\end{figure}
Combining the power corrections with the all-orders resummation we
obtain our final results for the jet shape at the Tevatron, shown in
Figs.~\ref{final50}--\ref{final250}.
Also shown for comparison are the original leading order results.  At
$E_T=50$~GeV, the combined effect of all the corrections we have
considered roughly doubles the amount of energy near the edge of the
jet.  Even at high transverse energy, $E_T=250$~GeV, it is increased by
about 50\%, although most of this is accounted for by running coupling
effects.

\section{The radial moment}

It was recently proposed that the first moment of the jet shape,
$\langle r \rangle$, is a good measure of perturbative QCD.  Since this
is simply the area under the curves of the previous section, we are in a
good position to calculate $\langle r \rangle$ and compare our results
with those of~[\ref{GGK3}].

We can again make the simple analytical approximation of the previous
section, and obtain
\begin{eqnarray}
  \langle r \rangle_q &=&
  \frac{C_F\as}{2\pi}R\left[ 8\log\Rsep+2+6/\Rsep-3\Rsep-2/\Rsep^2 \right]
  + \langle r \rangle_{ISR}, \\
  \langle r \rangle_g &=&
  \frac{C_A\as}{2\pi}R\left[
  8\log\Rsep+\smfrac23+8/\Rsep+\smfrac43/\Rsep^3-\smfrac25/\Rsep^4
  -4/\Rsep^2-\smfrac{14}5\Rsep \right]
\nonumber\\
  &+&
  \frac{T_RN_f\as}{2\pi}R\left[
  \smfrac83-4/\Rsep+4/\Rsep^2-\smfrac83/\Rsep^3+\smfrac45/\Rsep^4
  -\smfrac25\Rsep \right]
  + \langle r \rangle_{ISR}, \phantom{(99)}\\
  \langle r \rangle_{ISR} &=&
  \frac{C\as}{2\pi}R^3\left[
  4\Rsep^2-\smfrac43\Rsep^3-\smfrac43 \right].
\end{eqnarray}
Note that these are the same as in~[\ref{GGK3}], except that they
neglected the initial-state contribution\footnote{This comment applies
  to the analytical results with which they illustrate their argument.
  All their numerical results use the exact matrix elements so do
  correctly include this contribution.}.  For $R=1$, $\Rsep=1.3$, this
gives a contribution of 25\% to gluon jets and 45\% to quark jets, so is
clearly important in determining the $E_T$-dependence of the jet shape.

\begin{figure}
  \centerline{
    \resizebox{!}{6cm}{\includegraphics{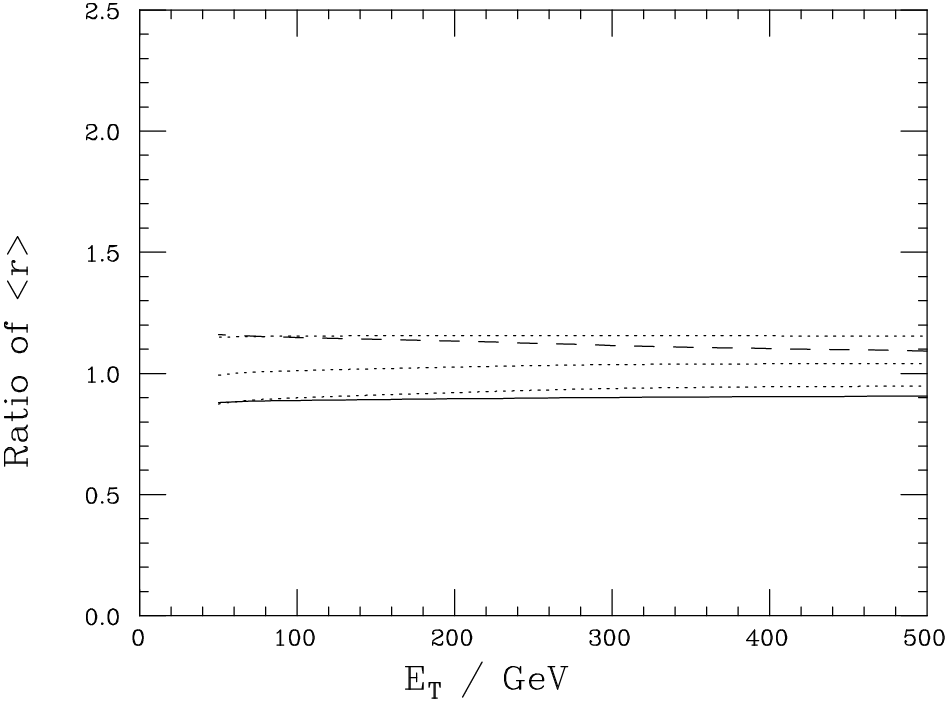}}
    }
  \caption[]{{\it Dependence of the radial moment on the factorization
      and renormalization scales, $\mu$.  Solid = ratio of
      $\mu\!\!=\!\!E_T$ to $\mu\!\!=\!\!E_T/2$, dashed =
      $\mu\!\!=\!\!E_T/4$ to $\mu\!\!=\!\!E_T/2$, with $R\!=\!0.7$,
      $\Rsep\!=\!2$, and CTEQ4M pdfs with $\alpha_s(M_Z)\!=\!0.116$.
      The dotted curves are the same but with the numerator evaluated
      using our analytical approximation.}}
  \label{mu}
\end{figure}

In Fig.~\ref{mu} we show the dependence of our results on the
renormalization and factorization scales.
We see that again the analytical approximation works very well, agreeing
with the exact matrix element calculation to within 5\% everywhere, and
considerably better at low $E_T$ where the data lie at present, so we use
it for the remainder of this section.  In this approximation, the only
dependence on the factorization scale is the changing mix of quark and
gluon jets.  Note that our scale-dependence is overall much smaller than
that of~[\ref{GGK3}], owing to the fact that we consistently truncate at
LO, whereas they include a partial set of NLO terms, as illustrated in
Eqs.~(\ref{wrong}) and~(\ref{right}).

Turning to the improvements we made for the jet shape distribution in
the previous section, we find that in the radial moment there are no
large logarithms to resum, so neither the running coupling nor the
Sudakov resummation make any formal difference to the results (although
it can be seen from Figs.~\ref{run} and~\ref{resum} that they do make
considerable numerical difference).  However, power corrections are
extremely important, as discussed in~[\ref{GGK3}].  Using our approach,
we can estimate the power corrections as
\begin{eqnarray}
  \langle r \rangle &\sim& \frac{4C_i}{2\pi} \int_0^R dr \int_0^1 dz
  \; \as(zrE_T) \\
  &=& \frac{4C_i}{2\pi} \int_0^{RE_T}
  \frac{dk}{E_T}\as(k)\log\frac{RE_T}{k},
\end{eqnarray}
and hence
\begin{eqnarray}
  \langle r \rangle_{\mathrm{pow}}
  &=& \frac{4C_i}{2\pi} \; \frac1{2\beta_0\as(RE_T)}
  \left(\frac{Q_0}{E_T}\right)
  \left(\abz(Q_0)-\as-2\beta_0\as^2\left(1+\log\frac\mu{Q_0}\right)\right).
\end{eqnarray}
From the initial-state contribution, we only find $1/E_T$ corrections
that are not enhanced by the extra logarithm, so we can neglect them.
Comparing our result with~[\ref{GGK3}], we find a big difference.  Their
power correction is $\sim1/\langle r \rangle_{LO}\sim1/C_i\as(E_T)$
while ours is $\sim C_i/\as(E_T)$.  Therefore their hadronization
corrections are smaller for gluon jets than quark jets, in contrast with
what one would expect from a simple colour tube model[\ref{W}], in which
a gluon jet is attached to two colour tubes and a quark jet is attached
to one.  Thus one would na\"\i vely expect twice the correction for a
gluon jet than for a quark jet.  This is in agreement with what we find,
$\langle r \rangle_{g\mathrm{pow}}/\langle r \rangle_{q\mathrm{pow}} =
C_A/C_F = 9/4$.  This can be expected to affect the $E_T$ dependence,
since the flavour mix changes considerably with $E_T$.

\begin{figure}
  \centerline{
    \resizebox{!}{6cm}{\includegraphics{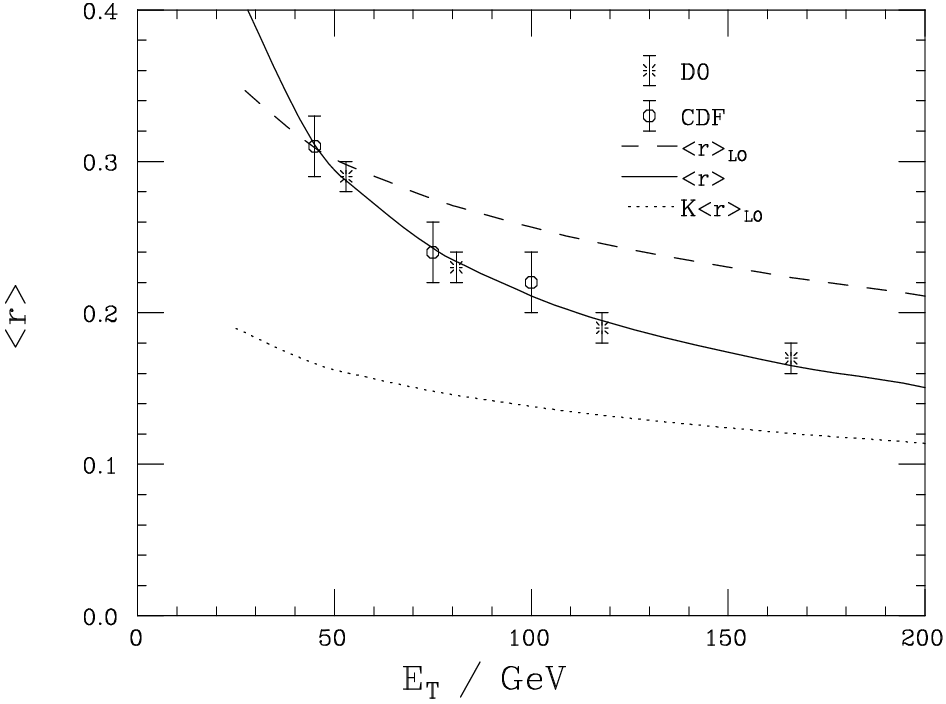}}
    }
  \caption[]{{\it The radial moment at the Tevatron.  Data are
      from~[\ref{CDF3},\ref{D03},\ref{GGK3}].  The dashed curve is the
      leading order perturbative result, with $R=1$, $\Rsep=1.3$ and
      CTEQ4M pdfs with $\alpha_s(M_Z)=0.116$.  The solid curve is our
      fit to the data.  The dotted curve is the perturbative component
      of the fit.}}
  \label{fit}
\end{figure}

We have repeated the fit to the data collated in~[\ref{GGK3}] using our
results.  Following their approach, we multiply the LO prediction by a
$K$ factor that is allowed to vary in the fit.  Fixing $\Rsep=1.3$ and
$\mu=E_T/2$, we obtain a good fit to the data, as shown in
Fig.~\ref{fit}.
Setting $Q_0=2$~GeV, we obtain $K=0.54\pm0.06$ and
$\abz(\mbox{2~GeV})=0.55\pm0.06$.  This is in very good agreement with
the results of~[\ref{DW1}], particularly bearing in mind that our
perturbative results are only LO while theirs were NLO.  We also get
good agreement for $Q_0=3$~GeV, with $K=0.52\pm0.06$ and
$\abz(\mbox{3~GeV})=0.44\pm0.04$.  In both cases the values of $K$ and
$\abz$ are strongly anticorrelated.

It is interesting to note that although our LO results are considerably
larger than those in~[\ref{GGK3}], presumably because of the partial set
of NLO terms they include, our $K$ factor comes out almost identical to
theirs.  This is presumably a reflection of the fact that the $E_T$
dependencies of the power-suppressed terms are considerably different
owing to the different flavour
dependencies.  If our perturbative contribution is forced to be as small
as theirs, the resulting $\abz$ does not agree with~[\ref{DW1}].

One might worry about how far from unity the fitted value of $K$ is.
However, we have already argued in the previous section that with
current jet definitions one expects large uncertainties and large higher
order corrections, so we are not unduly worried.  In fact in this
context it is worth pointing out that we can get an equally good quality
of fit, with comparable $\abz$ values, for any $\Rsep$ between 1 and 2.
Any change in $\Rsep$ is compensated by a change in $K$ in the fit.  It
is hard to draw any quantitative conclusions without improved jet
definitions and a full NLO calculation.

\section{Summary and discussion}

In this paper we have tried to examine all the important effects
associated with predicting the jet shape distribution in hadron
collisions.  In doing so, we are examining the details of how a hard
parton turns into a jet of hadrons.  We have estimated that large
corrections arise from high orders in perturbation theory and from
non-perturbative effects.  Traditionally, this has been regarded as a
reason to shy away from looking inside jets, but we would like to
advocate exactly the opposite view: the parton$\to$hadrons transition is
one of the most interesting outstanding questions in QCD, and it is
precisely because these effects are large that we should look inside
jets to study them.  Hadron collisions make a perfect place to make such
studies, because they are a source of high rates of jets over a very
wide range of jet scales in the same experiment.

However, we have also emphasized the importance of infrared safe jet
definitions.  These are essential, because in order to study the onset
of non-perturbative effects, we must have a perturbative `background'
that is well understood and under control.  If we use an infrared unsafe
jet definition, then the whole process is non-perturbative, and we have
no `solid ground' to start from.  Furthermore, if one wants to use the
jets for perturbative jet physics, for example to measure the inclusive
jet cross section, it is important that the definition be safe to all
orders, not just the order to which the theoretical calculations are
carried out.  With an infrared safe jet definition, non-perturbative
corrections are guaranteed to be suppressed by (at least) one power of
$\mbox{`1 GeV'}/E_T$, while if unsafe behaviour arises at any higher
order there are no such guarantees.  In particular, in an unsafe (or
`almost unsafe') algorithm, there is no reason to expect the cross
section to show the same scaling with energy as the perturbative
calculation.

We have critically examined the current norm in hadron collider jet
definitions: the iterative cone algorithm.  We have shown that it
becomes unsafe at next-to-next-to-leading order for jet cross sections,
which means next-to-leading order for internal jet properties.  Since
next-to-leading order is the minimum required for quantitative analysis,
this means that using current jet definitions one can never probe the
internal structure of a jet in a quantitative way.  We therefore
recommend that:
\begin{itemize}
\item Experiments should abandon cone-based algorithms, and use the
  version of the $\kt$ algorithm adapted to hadron
  collisions[\ref{CDSW},\ref{ES}].
\item If for some reason this is not possible, they should improve the
  cone-based algorithms as discussed in Sect.~\ref{Asolution}.  It would
  help ensure that the same definition is used in theory and experiment
  if the experiments made their jet algorithms publicly available.
\end{itemize}

We also discussed a `solution' that purported to cure these problems:
the introduction of an $\Rsep$ parameter.  In our view, this has
obscured the whole subject for several years, by giving the impression
that leading order perturbative QCD gave a good description of data on
the internal structure of jets, when we have absolutely no right to
expect it to.  We have shown that higher order corrections and
non-perturbative corrections can each be expected to be as large as 50\%
in some phase space regions, so it would be a miracle if the leading
order predictions gave a good fit to data.  At any given phase space
point, we can mask all our ignorance simply by adjusting $\Rsep$, and
only when we study the dependence on the jet kinematics do we get found
out.  This is illustrated in Fig.~\ref{fit} for example, where it is
clear that any one of the points could be fit by varying $\Rsep$
(i.e.~by shifting the dashed curve up or down by an arbitrary amount)
but the only ways to simultaneously fit all of them are either to
abandon all pretence that $\Rsep$ is related to the jet definition, and
just treat it as a fit parameter that is allowed to vary as an arbitrary
function of the jet kinematics, or to realize that there is more to the
internal structure of jets than leading order perturbative QCD.  It is
precisely what this `more' is, that we have tried to explore in this
paper.

In addition to the jet definition, we also discussed an ambiguity in the
jet shape definition.  In some experiments it is defined using all
particles (that are close enough to the jet axis) in the event, while in
others it is defined using only those particles that are assigned to the
jet by the jet definition.  For cone-type jet definitions there is
little to choose between them, but in the $\kt$ algorithm the
next-to-leading order corrections are much better behaved in the latter
case than the former (see Sect.~\ref{higher}).  We therefore recommend
that:
\begin{itemize}
\item Experiments define the jet shape distribution using only those
  particles assigned to the jet, and no other particles that happen to
  be nearby.
\end{itemize}

It is hoped that full next-to-leading order calculations will soon be
available for the jet shape, enabling quantitative analyses to be made
for the first time.  Next-to-leading order has the advantage of
modelling effects that happen near edge of the jet~-- merging and
splitting issues in cone algorithms and the raising of the `kinematic
zero' at $r\!=\!R$ in the $\kt$ algorithm.  It also has a much better
defined normalization, so the difference between the data and the
predictions will provide a strong constraint on non-perturbative
corrections.

We have made an approximate calculation of the next-to-leading order
corrections.  Although it is too crude to predict the expected
normalization, it should provide a good model of the kinematic effects.
We therefore predict that when the full calculation is made, the
following features will be found:
\begin{itemize}
\item The corrections in the iterative cone algorithm will be infrared
  unsafe (in the method of~[\ref{GK}] this will manifest itself as an
  $s_{\mathrm{min}}$ dependence analogous to our $E_0$ dependence in
  Fig.~\ref{itshape}).
\item In the improved cone algorithm, the corrections will be large and
  negative, diverging to negative infinity at both small and large $r$.
\item In the $\kt$ algorithm, the corrections will also diverge to
  negative infinity at small~$r$.  At large $r$ the results will depend
  critically on whether all particles are used (in which case they will
  diverge to positive infinity), or only those particles in the jet (in
  which case they will have a small cusp at $r\!=\!R$).
\end{itemize}

Once the next-to-leading order calculation is available, by far the
largest missing effects will be (at small $r$) higher order terms
enhanced by logarithms of $r$ and (at medium to large $r$)
non-perturbative power-suppressed corrections.  In this paper we have
shown how to resum the large logarithms to modified leading logarithmic
accuracy (i.e.~summing terms like $\as^n\log^{2n-1}r$ and
$\as^n\log^{2n-2}r$ to all orders in $\as$), and we expect that this can
be extended to a full next-to-leading logarithmic resummation.  We have
also calculated in the Dokshitzer-Webber approach the effect of
power-suppressed terms on the jet shape distribution.  Although these
depend on fundamentally non-perturbative parameters, in their approach
these parameters are postulated to be universal, and have already been
measured in $e^+e^-$ annihilation.

It is important to note that both of these effects are independent of
the precise details of the jet definition (in the $\Rsep$ language, they
are completely $\Rsep$-independent).  Therefore it would be
straightforward to combine them with the next-to-leading order
calculation, when available.

One source of power-suppressed terms that we have not discussed is the
`underlying event', the soft hadrons that are sprayed around the event
by secondary interactions between the hadron remnants.  As estimated
in~[\ref{GGK3},\ref{So}], this is less important than the hadronization
corrections that we have considered in this paper.  Nevertheless, since
it has the same energy-dependence as the hadronization corrections, it
cannot easily be separated out, and at some level it must be understood
before fine details of the hadronization can be extracted.  Jet shape
analyses in photoproduction could prove crucial from this point of view
since one can `tune' the underlying event in or out by selecting
different dijet kinematics.  Unfortunately however, it seems that
next-to-leading order calculations for photoproduction are more
distant than for hadron collisions.

In Ref.~[\ref{GGK3}], the first moment of the jet shape distribution was
proposed as a good measure of jet shape.  From the perturbative point of
view, it has the disadvantage that power corrections are larger in the
radial moment ($\sim1/\as E_T$) than in the full distribution
($\sim1/rE_T$).  One advantage of the radial moment is that it is a
single quantity that can be plotted as a function of the jet kinematics,
rather than a whole distribution.  It is also free from large
logarithmic corrections.  However, both these statements can be equally
applied to some measure such as the integral of $r\psi(r)$ from 0.25 to
0.75, which has power corrections that are smaller by a factor $\as$.
In our opinion, either approach throws away important dynamic
information that is contained in the shape of the distribution, and one
would be much better off simply measuring $\psi(r)$ over as wide a range
of $r$ and event kinematics as possible, and then possibly neglecting
the extreme regions of small and large $r$ in the comparison with
theory, exactly as $e^+e^-$ annihilation experiments do with event
shapes.

Despite our misgivings about the quality of the jet definition, and the
use of only a leading order calculation, we made a fit to the
experimental data on the radial moment.  It is encouraging that the
resulting non-perturbative parameter is in agreement with that extracted
from $e^+e^-$ annihilation data.  If confirmed with a full
next-to-leading order perturbative calculation, this would be a highly
non-trivial test of the universality implicit in the Dokshitzer-Webber
approach.  In particular, these coefficients have never been measured in
gluon-induced jets.

There has been a rapidly-increasing interest in the internal structure
of jets in recent years, which is now being spurred on by the promise
that the next-to-leading order corrections are imminent.  We believe
that jet studies in hadron colliders are at a similar point to $e^+e^-$
annihilation before the Z factories, and that we are about to enter a
new era of high precision predictions and measurements of a wide range
of jet properties.  To realize this potential we need close
collaboration between theorists and experimenters to define observables
that are sensitive to the interesting physics and feasible to calculate,
and measure, accurately.  The jet shape distribution considered in
this paper is only scratching the surface.

\newpage
\section*{References}

\begin{enumerate}
\item\label{CDF1}
  F. Abe {\it et al.}, CDF Collaboration,
  Phys.\ Rev.\ Lett.\ 77 (1996) 438.
\item\label{D01}
  G.C. Blazey, for the D\O\ Collaboration,
  in Proceedings of the 31st Rencontres de Moriond: QCD and High-energy
  Hadronic Interactions, Les Arcs, France, 1996, p.~155.
\item\label{CDF2}
  F. Abe {\it et al.}, CDF Collaboration,
  Phys.\ Rev.\ Lett.\ 77 (1996) 5336, erratum {\it ibid.} 78 (1997) 4307;
  Phys.\ Rev.\ D54 (1996) 4221.
\item\label{D02}
  S. Abachi {\it et al.}, D\O\ Collaboration,
  Phys.\ Rev.\ D53 (1996) 6000.
\item\label{ACGG}
  F. Aversa, P. Chiappetta, M. Greco and J.P. Guillet,
  Phys.\ Rev.\ Lett.\ 65 (1990) 401;
  Z.\ Phys.\ C46 (1990) 253;\\
  F. Aversa, P. Chiappetta, L. Gonzales, M. Greco and J.P. Guillet,
  Z.\ Phys.\ C49 (1991) 459.
\item\label{EKS1}
  S.D. Ellis, Z. Kunszt and D.E. Soper,
  Phys.\ Rev.\ Lett.\ 62 (1989) 726;
  Phys.\ Rev.\ D40 (1989) 2188;
  Phys.\ Rev.\ Lett.\ 64 (1990) 2121.
\item\label{GGK1}
  W.T. Giele, E.W.N. Glover and D.A. Kosower,
  Nucl.\ Phys.\ B403 (1993) 633;
  Phys.\ Rev.\ Lett.\ 73 (1994) 2019.
\item\label{EKS2}
  S.D. Ellis, Z. Kunszt and D.E. Soper,
  Phys.\ Rev.\ Lett.\ 69 (1992) 1496;\\
  S.D. Ellis and D.E. Soper,
  Phys.\ Rev.\ Lett.\ 74 (1995) 5182.
\item\label{GGK2}
  W.T. Giele, E.W.N. Glover and D.A. Kosower,
  Phys.\ Rev.\ D52 (1995) 1486.
\item\label{GK}
  W.T. Giele and W.B. Kilgore,
  Phys.\ Rev.\ D55 (1997) 7183;\\
  W.B. Kilgore, `Next-to-leading Order Three Jet Production At Hadron
  Colliders', hep-ph/9705384.
\item\label{CDF3}
  F. Abe {\it et al.}, CDF Collaboration,
  Phys.\ Rev.\ Lett.\ 70 (1993) 713.
\item\label{D03}
  S. Abachi {\it et al.}, D\O\ Collaboration,
  Phys.\ Lett.\ B357 (1995) 500.
\item\label{ZEUS}
  M. Derrick {\it et al.}, ZEUS Collaboration, contribution to the
  XXVIII ICHEP, Warsaw, 1996, pa 02-043; DESY preprint in preparation.
\item\label{EKS3}
  S.D. Ellis, Z. Kunszt and D.E. Soper,
  Phys.\ Rev.\ Lett.\ 69 (1992) 3615.
\item\label{GGK3}
  W.T. Giele, E.W.N. Glover and D.A. Kosower,
  `Jet Investigations Using The Radial Moment', hep-ph/9706210.
\item\label{CDSW}
  S. Catani, Yu.L. Dokshitzer, M.H. Seymour, B.R. Webber,
  Nucl.\ Phys.\ B406 (1993) 187.
\item\label{CDF4}
  F. Abe {\it et al.}, CDF Collaboration,
  Phys.\ Rev.\ D45 (1992) 1448.
\item\label{LdP}
  S.D. Ellis, private communication to the OPAL Collaboration;\\
  D.E. Soper and H.-C. Yang, private communication to the OPAL
  Collaboration;\\
  L.A. del Pozo, University of Cambridge PhD thesis, RALT--002, 1993;\\
  R. Akers {\it et al.}, OPAL Collaboration,
  Z.\ Phys.\ C63 (1994) 197.
\item\label{D04}
  B. Abbott, M. Bhattarcharjee, D. Elvira, F. Nang and H. Weerts,
  `Fixed Cone Jet Definitions in D\O\ and $\Rsep$'
  Fermilab--Pub--97/242--E.
\item\label{Snowmass}
  J.E. Huth {\it et al.}, in {\it Research Directions for the Decade},
  Proceedings of the Summer Study on High Energy Physics, Snowmass,
  Colorado, 1990, p.~134.
\item\label{ES}
  S.D. Ellis and D.E. Soper,
  Phys.\ Rev.\ D48 (1993) 3160.
\item\label{KK}
  M. Klasen and G. Kramer,
  `Jet Shapes in $ep$ and $p\bar{p}$ Collisions in NLO QCD',
  hep-ph/9701247.
\item\label{K}
  H. Kuijf, `NJETS: A Monte Carlo for QCD Backgrounds', unpublished;\\
  F.A. Berends and H. Kuijf,
  Nucl.\ Phys.\ B353 (1991) 59.
\item\label{bible}
  Yu.L. Dokshitzer, V.A. Khoze, A.H. Mueller and S.I. Troyan,
  {\it Basics of Perturbative QCD}\/, Editions Fronti\`eres,
  Paris, 1991.
\item\label{S}
  M.H. Seymour,
  Nucl.\ Phys.\ B421 (1994) 545.
\item\label{CTEQ}
  H.L. Lai {\it et al.}, CTEQ Collaboration,
  Phys.\ Rev.\ D55 (1997) 1280.
\item\label{CW}
  S. Catani and B.R. Webber, University of Cambridge preprint
  Cavendish--HEP--97/10, in preparation.
\item\label{DW1}
  Yu.L. Dokshitzer and B.R. Webber,
  Phys.\ Lett.\ B352 (1995) 451.
\item\label{T}
  G.\ Turnock, `Energy-Energy Correlation Distribution in $e^+e^-$
  Annihilation', University of Cambridge preprint Cavendish--HEP--92/3,
  unpublished.
\item\label{RW}
  P.E.L. Rakow and B.R. Webber,
  Nucl.\ Phys.\ B187 (1981) 254.
\item\label{DW2}
  Yu.L. Dokshitzer and B.R. Webber, `Power Corrections to Event Shape
  Distributions', hep-ph/9704298.
\item\label{W}
  B.R. Webber, `Hadronization', lectures given at Summer School on {\it
  Hadronic Aspects of Collider Physics}, Zuoz, Switzerland, 1994,
  hep-ph/9411384.
\item\label{So}
  D.E. Soper, `Jet Observables in Theory and Reality', hep-ph/9706320.
\end{enumerate}

\end{document}